\documentclass[aps,
twocolumn,pra]{revtex4}
\usepackage{latexsym}
\usepackage{amsmath}
\usepackage{mathrsfs}
\usepackage{amsthm}
\usepackage{amssymb}
\usepackage{amsfonts}
\usepackage{fancybox}
\usepackage{mathbbol}
\usepackage{bm}
\usepackage{fancyhdr}
\usepackage{subfigure}
\usepackage{graphicx}
\usepackage{color}

\newcommand{\PT}{{\cal PT}}
\newcommand{\p}{{\cal P}}
\newcommand{\tp}{{\cal T}}
\newcommand{\bw}{{\bf w}}
\newcommand{\bW}{{\bf W}}
\newcommand{\tbw}{\tilde{\bf w}}
\newcommand{\tE}{\tilde{E}}
\newcommand{\vep}{{\varepsilon}}

\begin{document}
\everymath{\displaystyle}

\title{$\PT$-symmetric coupler with $\chi^{(2)}$ nonlinearity}

\author{K. Li$^*$, D. A. Zezyulin$^{\dag}$, P. G. Kevrekidis$^*$, V. V. Konotop$^{\dag}$, and F. Kh. Abdullaev$^{\ddag,\flat}$}

\affiliation{$^*$ Department of Mathematics and Statistics, University of Massachusetts,
Amherst, Massachusetts 01003-4515, USA\\
 $^\dag$Centro de F\'{\i}sica Te\'orica e Computacional, and Departamento de F\'{\i}sica, Faculdade de Ci\^encias,
 Universidade de Lisboa,
Avenida Professor Gama
Pinto 2, Lisboa 1649-003, Portugal\\
$^\ddag$ Instituto de F{\'i}sica T{\'e}orica, Universidade Estadual Paulista, 01140-070, Sao Paulo, Sao Paulo, Brazil\\
$^\flat$ Department of Physics, Kulliyyah of Science, International Islamic University of Malaysia, Jalan
Istana, Bandar Indera Mahkota 25200, Kuantan, Malaysia
}

\date{\today}

\begin{abstract}
We introduce the notion of a $\PT$-symmetric
dimer with a $\chi^{(2)}$ nonlinearity. Similarly to the Kerr case, we
argue that such a nonlinearity should be accessible in a pair of
optical waveguides with quadratic nonlinearity and gain and loss,
respectively. An interesting feature of the problem is that because
of the two harmonics, there exist in general two distinct gain/loss
parameters, different values of which are considered herein. We
find a number of traits that appear to be absent in the more
standard cubic case.
For instance, bifurcations of nonlinear modes from the linear solutions occur in two different ways depending on whether the first or the second harmonic amplitude is vanishing in the underlying linear eigenvector.
Moreover, a host of interesting bifurcation phenomena
appear to occur including saddle-center and   pitchfork bifurcations which our parametric variations
elucidate. The existence and stability analysis of the stationary
solutions is corroborated by numerical time-evolution simulations
exploring the evolution of the different configurations, when unstable.
\end{abstract}

\maketitle
\section{Introduction}

In the past fifteen years, the remarkable original proposal
of Refs.~\cite{R1}, relaying a potential physical relevance to Hamiltonians
respecting parity ($\p$) and time-reversal ($\tp$) symmetries,
has received considerable attention~\cite{R2}.
This proposal has highlighted the interest in considering (as operators potentially
bearing real spectra) Hamiltonians that are invariant
under these fundamental symmetries as an extension of the more
standardly considered self-adjoint Hamiltonian operators of quantum mechanics.
While for a decade since their inception, these notions were studied intensely at the linear
level, especially in the mathematical community (see e.g.
the review of~\cite{R2}), more recently it was realized
that linear optics~\cite{Muga} could present an ideal
playground for the realization of such non-Hermitian settings
(i.e., in ``open'' systems bearing gain and loss but in a $\PT$-symmetric form).
In particular, discrete systems~\cite{Kulishov,dimer_OL} with balanced gain and loss have been suggested as reduced models of the non-Hermitian optics,  obeying remarkable properties of waveguiding  and giving origin to the
blossoming field of discrete $\PT$-symmetric optics. $\PT$ symmetry was
thus first studied experimentally in the optical experiments of~\cite{salamo,dncnat,scherer,pertsch}. As a natural extension of optical applications,  it was suggested in Ref.~\cite{ziad} to consider nonlinear optical systems whose linear limit is $\PT$ symmetric, and in particular it  was shown that such systems with a periodic potential support stable solitons.

The simplest basic element of the discrete $\PT$-symmetric optics is a dimer with one site subjected to dissipation and another site subjected to gain~\cite{Kulishov,dimer_OL}. As a natural application of this system, one can consider a coupler with one active and one lossy waveguide~\cite{dncnat}. When the nonlinear effects are included, one deals with a $\PT$-symmetric nonlinear coupler, i.e. mathematically with a $\PT$-symmetric nonlinear dimer. Such a dimer with
a Kerr-type nonlinearity was intensively studied showing remarkable properties. In particular, it was shown in ~\cite{Ramezani} that such a nonlinear coupler is an integrable system allowing for a solution in the form of
quadratures.
The effect of nonlinear suppression of the periodic time reversals   and the
beam switching to the waveguide with gain was reported in~\cite{sukh1}
(a similar switching effect can also be implemented with
$\PT$-symmetric impurities inserted in an otherwise conservative
coupler~\cite{R30add3}).

Further studies of the nonlinear discrete optical systems were performed in a
number of directions. We mention a few of these in what follows.
First, including one more coordinate (in addition to the evolution one; in optical applications such systems could be seen as coupled {planar} waveguides with balanced gain and loss) made it relevant to consider the dynamics of bright~\cite{bright} and dark~\cite{dark} solitons, breathers~\cite{breathers}, as well as instabilities and rogue waves~\cite{rogue}. Another extension of nonlinear dimer activity is related to the inclusion of $\PT$-symmetric defects in discrete nonlinear systems. In the latter context, problems such as nonlinear wave
scattering~\cite{dmitriev1} (see also~\cite{jennie})
and the lifting of the degeneracy of
discrete vortices~\cite{leykam} were considered. Finally, a nonlinear dimer or more generally the so-called nonlinear ``oligomers'' (i.e., few site configurations) introduced in~\cite{pgk} (see also~\cite{uwe}), were shown to
allow for the existence of continuous families of nonlinear modes~\cite{konorecent3}. Among these, a $\PT$-symmetric quadrimer model naturally appears in the description of light propagation in a birefringent coupler~\cite{LZKK}.  Discrete solitons in different types of infinite $\PT$-symmetric waveguide arrays
were studied numerically~\cite{dmitriev2} and analytical proofs for their
existence have been proposed using the anticontnuum limit~\cite{KPZ} and
via analysis of the modes bifurcating from the linear limit~\cite{tyugin}. Solitons in a necklace of coupled   dispersive  waveguides were reported in \cite{necklace}.

Most of the above investigations have taken place at the level of
the well-known Kerr-type nonlinearity.  Existence of nonlinear modes and integrals of motion in $\PT$-symmetric systems with  more
general cubic nonlinearities  was investigated in~\cite{ZK_nonlin}.
Nevertheless, another type
of nonlinearity of particular interest to optics is the quadratic
one~\cite{buryak}. In the latter context,
switching in two parallel waveguides was studied in
\cite{Schiek94}. The  intensity-dependent switching in lithium niobate directional couplers was subsequently first observed in \cite{Schiek96}.
 The particular case of a dimer with one nonlinear and one linear waveguide
was considered in \cite{Assanto}.
Furthermore, nonlinear localized  modes in  arrays with a quadratic
nonlinearity are considered in numerous
works~\cite{Mak97,Peschel98,Miller,Darmanyan}.
Experimentally~\cite{Pertsch}, the fundamental modes of a second harmonic are
strongly  confined, so the coupling constant between second harmonic modes in
different waveguides
is very small and sometimes  can even
be neglected. Discrete solitary waves in this configuration
were systematically probed in~\cite{Iwanow}. The plane waves and
localized modes
in this case were considered in~\cite{Kevrekidis}. Extensions in the
case of two-dimensional states including discrete vortices were
proposed in~\cite{pgk2}, while the mobility of the solitary waves
in both one and two dimensions was explored in~\cite{pgk3}.

Recently the studies of solitons in quadratically nonlinear
media were extended to $\PT$-symmetric systems.
More specifically, the existence and stability of solitons for
localized potentials in quadratic media
was explored in~\cite{Mor1}. In the work of~ \cite{Mor2},
the effect of periodic  $\PT$-symmetric potentials on $\chi^{(2)}$ solitons was described.

It is on that direction of exploring the interplay of
quadratic nonlinearity and $\PT$-symmetric potentials that
the present work is focused. It is appreciated that even
 {in the   case of two waveguides,}
this combination offers
a significant level of complexity, as well as a number of
features that are absent in the cubic Kerr nonlinearity case.
In particular,  due to quadratic nonlinearity and the  particular structure of  eigenvectors of the  underlying linear $\PT$-symmetric operator, continuation of the linear eigenvectors into the nonlinear domain is performed in two different ways, depending on whether the first or the second harmonic is vanishing in the linear eigenvector. We develop perturbative formal expansions that enable us
to capture  analytically these  two different types of bifurcations  of the nonlinear modes from the linear solutions. Next, we employ numerical computations and  observe the symmetry
breaking, as well as saddle-node bifurcations and
identify the stability characteristics
of the
solutions. In the case of instability,
the dynamical evolution of the configurations is explored.

Our presentation is structured as follows. In section~\ref{linear}, upon
introducing the general model, we focus on
its linear analysis. In section~\ref{sec:bif}, we introduce
nonlinearity and analytically
explore how it affects the linear  modes
(i.e., consider the bifurcations of nonlinear modes from linear ones).
In section~\ref{sec:nonlin}, we corroborate our analytical considerations with
detailed numerical results identifying    the nonlinear modes and
their stability for different values of
the propagation constant
and the gain/loss strength parameters. For the unstable modes,
we touch upon their dynamical evolution in section~\ref{sec:dyn}. Finally,
in section~\ref{sec:concl}, we summarize our findings and present some
interesting directions for future studies.

\section{The Model and its linear Analysis}
\label{linear}%

The prototypical setup of equations describing the $\PT$ symmetric coupler with  quadratic nonlinearity
reads as follows:
\begin{subequations}
\label{basic_equations}
\begin{eqnarray}
i\dot{u}_{1} &=& k_1 u_2-2u_1^* v_1 + i\gamma_1 u_1,\\
i\dot{v}_{1} &=& k_2 v_2-u_1^2-q v_1 + i\gamma_2 v_1,\\
i\dot{u}_{2} &=& k_1 u_1-2u_2^* v_2 - i\gamma_1 u_2,\\
i\dot{v}_{2} &=& k_2 v_1-u_2^2-q v_2 - i\gamma_2 v_2.
\end{eqnarray}
\end{subequations}
Each waveguide contains two harmonics: the fundamental field (first harmonic)  $u_j$ and the second harmonic $v_j$, $j=1,2$, which are nonlinearly coupled. The linear coupling
between the first harmonics is characterized by the parameter $k_1$, while that of the second harmonics by
$k_2$. Both $k_1$ and $k_2$ will be considered positive. The gain (loss) strength
in the two arms of the dimer is given by the parameters $\gamma_{j}>0$ ($\gamma_{j}<0$), for the first ($j=1$) and second ($j=2$) harmonics, respectively.
In what follows, we will explore different parameter values of
$(\gamma_1,\gamma_2)$ to get a systematic sense of the model phenomenology.
The overdot in (\ref{basic_equations}) denotes the derivative with respect to the evolution
variable which, here, we will denote as $t$ (although in the optical
realm it represents the propagation distance $z$).

Being interested in the stationary modes, we make use of the ansatz
\begin{eqnarray}
\label{lin_ansatz}
\left(
\begin{array}{c}
u_1 \\ v_1 \\u_2 \\ v_2
\end{array}
\right)=e^{-i\Lambda Et} \bw, \quad\mbox{where}\quad \Lambda=\left(
\begin{array}{cccc}
1&0&0&0
\\
0&2&0&0\\
0&0&1&0
\\ 0&0&0&2
\end{array}
\right),
\end{eqnarray}
 $E$ is the propagation constant, and $\bw$ is $4\times 1$ constant column vector. This reduces (\ref{basic_equations}) to the eigenvalue problem
\begin{equation}
\label{eq:stat}
E\Lambda\bw = H\bw + F(\bw) \bw,
\end{equation}
with the  respective linear
operator given by
\begin{equation}
\label{eq:H} {H}=\left(\!
         \begin{array}{cccc}
           i\gamma_1 & 0 & k_1 & 0\\
           0 & i\gamma_2 - q & 0 & k_2\\
           k_1 & 0 & -i\gamma_1 & 0 \\
           0 & k2 & 0 & -i\gamma_2 -q
         \end{array}
       \!\right)
\end{equation}
and the nonlinear part described by the matrix-function
\begin{equation}
\label{F}
F(\bw)=-\left(\!\!
         \begin{array}{cccc}
           0 & 2(w^{(1)})^* & 0 & 0\\
           w^{(1)} & 0 & 0 & 0\\
           0 & 0 & 0 & 2(w^{(3)})^* \\
           0 & 0 & w^{(3)} & 0
         \end{array}
       \!\right),
\end{equation}
where $w^{(j)}$ are used for the entries of vector $\bw$.

It is easy to check that $H$ is $\PT$-symmetric with respect to the action of the
parity operator
\begin{eqnarray}
\label{eq:P} {\p}=\left(\!\!
         \begin{array}{cccc}
           0 & 0 & 1 & 0\\
           0 & 0 & 0 & 1\\
           1 & 0 & 0 & 0 \\
           0 & 1 & 0 & 0
         \end{array}
       \!\right),
\end{eqnarray}
and time reversal operator ${\cal T}$ performing the complex
conjugation (along with $t \rightarrow -t$): $H\PT = \PT H$.

In the linear case [when one neglects  all the
nonlinear terms, i.e $F(\bw)\bw$] the problem decouples into two $\PT$-symmetric
dimers: the first one is composed of fields $u_1$ and $u_2$ and
another one of the fields $v_1$ and $v_2$. Now the eigenvalue problem (\ref{eq:stat}) is reduced to
$H\tbw=\Lambda\tE\tbw$, where we use tilde in order to refer to eigenvectors and eigenvalues of the linear problem. Solutions of the latter equation can be found from the spectrum of the operator $\Lambda^{-1}H$.  The computation yields the following linear eigenvalues
\begin{eqnarray}
\label{E} \tE_{1,2}=\pm\sqrt{k_1^2-\gamma_1^2}, \quad \tE_{3, 4}
=\frac12\left(-q\pm\sqrt{k_2^2-\gamma_2^2}\right).
\end{eqnarray}
The eigenvectors associated with the eigenvalues $\tE_{1,2}$  can be
written down as follows:
\begin{eqnarray}
\label{eq:w12} \tbw_1= \left(\!\!
         \begin{array}{c}
           e^{i\theta_1/2}\\
           0\\
           e^{-i\theta_1/2}\\
           0
         \end{array}
       \!\right), \quad \tbw_2= \left(\!\!
         \begin{array}{c}
           ie^{-i\theta_1/2}\\
           0\\
           -ie^{i\theta_1/2}\\
           0
         \end{array}
       \!\right),
\end{eqnarray}
while for $\tE_{3,4}$ one has
\begin{eqnarray}
\label{eq:w34} \tbw_3= \left(\!\!
         \begin{array}{c}
          0\\
            e^{i\theta_2/2}\\
          0\\
            e^{-i\theta_2/2}
         \end{array}
       \!\right), \quad \tbw_4= \left(\!\!
         \begin{array}{c}
           0\\
           ie^{-i\theta_2/2}\\
           0\\
           -ie^{i\theta_2/2}
         \end{array}
       \!\right),
\end{eqnarray}
where $\theta_{1,2} = \arctan\left(
\frac{\gamma_{1,2}}{\sqrt{k_{1,2}^2 - \gamma_{1,2}^2}}\right)$. Introducing the inner product as $\langle \bw_1, \bw_2 \rangle = \bw_1^\dag \bw_2$ (hereafter $\bw^\dag = (\bw^T)^*$ is the Hermitian conjugation), we observe that the  linear eigenvectors
obey the following relation:
\begin{equation}
\langle \tbw_j^*,\Lambda\tbw_p\rangle=\tbw_j^T\Lambda\tbw_p=0, \quad j\neq p.
\label{ortt}
\end{equation}
Notice however that  $\langle \tbw_j^*,\Lambda\tbw_j\rangle \ne 0$.

Generally speaking, the existence of two different gain/loss coefficients in $\PT$-symmetric lattices  results in the existence of different ``phases'' \cite{konorecent3,ZK_nonlin} featuring different linear properties of the model. In the case at hand the linear part of the system can belong to one of the
{  four} phases: (i) unbroken (or \textit{exact}) $\PT$
symmetry when all the four eigenvalues  (\ref{E}) are real; this phase
corresponds to the rectangle given by the inequalities   $|\gamma_{1}| < k_{1}$ and $|\gamma_{2}| < k_{2}$; (ii)
 $\tE_{1,2}$ are complex conjugates while   $\tE_{3,4}$ remain   real
which corresponds to $|\gamma_1|>k_1$ and  $|\gamma_2|<k_2$;
 (iii) \textit{vice versa},
$\tE_{1,2}$ are real while $\tE_{3,4}$ are complex conjugates; this
corresponds to   $|\gamma_1|<k_1$  and $|\gamma_2|>k_2$;
 (iv)
all four  eigenvalues have nonzero imaginary part, i.e.
$|\gamma_{1}|>k_{1}$ and $|\gamma_{2}|>k_{2}$. Phases (ii)-(iv) correspond to the
broken $\PT$ symmetry. On the plane ($\gamma_1,\gamma_2)$ there exist four \textit{quadruple points}
corresponding to the corners of the above mentioned rectangle: $(\gamma_1,\gamma_2)=(\pm k_1, \pm k_2)$. These are the exceptional points  where all
four phases touch.

If the $\PT$ symmetry is unbroken, i.e.  $|\gamma_{1}| < k_{1}$ and $|\gamma_{2}| < k_{2}$, then the choice of
the eigenvectors in Eqs.~(\ref{eq:w12})--(\ref{eq:w34})  makes
them    $\PT$ {\em invariant}, i.e. $\PT \tbw = \tbw$. One can see that in the absence of the degeneracy, i.e. at all $\tE_j$ different, in
the stationary linear regime the total energy is concentrated in only one
harmonic of each waveguide. Namely, the field is guided only in the first (second) harmonic, i.e.  $v_{1,2}=0$ ($u_{1,2}=0$), for the
eigenvectors corresponding to the eigenvalues $\tE_{1,2}$ ($\tE_{3,4}$).
This will prove rather critical in some of the considerations
that follow
 (especially as regards the continuation of nonlinear
modes from the linear limit).
Notice that the linear modes $\tbw_3$ and $\tbw_4$  at the same time solve the full (i.e. the nonlinear) system (\ref{basic_equations}) because in the absence of the energy guided in the fundamental mode, the system is effectively linear: $F(\tbw_{3,4})=0$.

\section{Bifurcations of nonlinear modes from the linear eigenstates}
\label{sec:bif}
In the previous section we computed solutions of the linear problem which can be formally obtained from the full nonlinear problem (\ref{eq:stat})  by neglecting the nonlinear term $F(\bw)\bw$. In this section we look for solutions of the full  nonlinear problem bifurcating from the linear solutions. To this end  we construct formal small parameter expansions around the linear  eigenvectors (\ref{eq:w12}) and (\ref{eq:w34}).  However, properties of the eigenvectors $\tbw_{1,2}$ and $\tbw_{3,4}$ are essentially different:  the latter couple of linear eigenvectors simultaneously solve the full nonlinear problem  because the nonlinear operator vanishes at them, i.e., $F(\bw_{3,4})=0$. However for the eigenvectors $\tbw_{1,2}$ one has  $F(\bw_{1,2})\ne 0$. This suggests that the formal expansions for the
nonlinear modes bifurcating from   $\tbw_{1,2}$ and $\tbw_{3,4}$ should be constructed in different ways.

\subsection{Nonlinear modes bifurcating from $\tbw_{1,2}$}

Since  $F(\bw_{1,2})\ne 0$, one can expect that the linear solutions  $\tbw_{1,2}$ can approximate nonlinear modes only in a situation when the nonlinear term $F(\bw)\bw$  is negligible in Eq.~(\ref{eq:stat}).  This  situation takes place if one considers  nonlinear modes $\bw$ of small amplitude, i.e. at $\|\bw\| \to 0$ which is usually referred to as the {\em linear limit}  (here $\|\tbw\|$ stands for a norm of the vector $\tbw$, which could be, say, the Euclidean one). Therefore, let us search  for small-amplitude nonlinear modes bifurcating from the linear solutions $\tbw_{1,2}$. We assume that $\PT$ symmetry is unbroken and all the eigenvalues (\ref{E}) are distinct from each other  and  introduce  the following formal expansions for the nonlinear modes $\bw_j$ bifurcating from the $j$th linear eigenstate ($j=1,2$):
\begin{eqnarray}
\label{expan_w}
\bw_j  = \vep\tbw_j + \vep^2 \bW_j^{(1)} + \vep^3\bW_j^{(2)}+\cdots,\\
\label{expan_E}
E_j  =   \tE_j + \vep e_j^{(1)} +  \vep^2 e_j^{(2)}+\cdots.
\end{eqnarray}
Here $\vep$ is a small positive  formal parameter, and    $\bW_j^{(1,2,\ldots)}$ and $e_j^{(1,2,\ldots)}$ are the vectors and the coefficients to be determined.
The expansion (\ref{expan_w}) and the definition (\ref{F}) imply that $F(\bw_j) = \vep F(\tbw_j) + \vep^2 F(\bW_j^{(1)}) + \cdots$.

 Since the linear eigenvectors $\tbw_p$ ($p=1,2,3,4$) constitute a complete basis, we can search  for the correction  $\bW_j^{(1)}$ in the form
\begin{eqnarray}
\label{exp_W}
\bW_j^{(1)}=\sum_{p=1, \, p\neq j}^4 c_p \tbw_p
\end{eqnarray}
(notice that in the latter equation we set $c_j=0$ which can always be achieved by means of renormalization of the small parameter $\vep$).
Substituting the introduced expansions into the nonlinear problem (\ref{eq:stat}), in the $\vep^2$-order we have
\begin{equation}
\label{ord2}
\tE_j\Lambda\bW_j^{(1)}+e_j^{(1)}\Lambda\tbw_j = H\bW_j^{(1)} + F(\tbw_j)\tbw_j.
\end{equation}

An unusual property of the case at hand is that the nonlinearity is orthogonal to the states $\tbw_j^*$ and $\tbw_{3-j}^*$:
\begin{eqnarray}
\langle\tbw_j^*,F(\tbw_j)\tbw_j\rangle=\langle\tbw_{3-j}^*,F(\tbw_j)\tbw_j\rangle=0.
\end{eqnarray}
Thus applying $\tbw_j^\dag$ and $\tbw_{3-j}^\dag$ to the both sides of (\ref{ord2}), using that $H^\dag=H^*$ and $\Lambda^\dag=\Lambda$ and accounting for  (\ref{ortt}), one readily finds that $e_j^{(1)}=0$ and $c_{3-j}=0$. Finally, applying $\tbw_3^\dag$ and $\tbw_4^\dag$, we compute the coefficients $c_{3,4}$ explicitly which yields the following expression:
\begin{equation}
\label{W1}
\bW_j^{(1)}=\frac{\langle\tbw_3^*,F(\tbw_j)\tbw_j\rangle\tbw_3}{(\tE_j-\tE_3)\langle\tbw_3^*,\Lambda\tbw_3\rangle} +\frac{\langle\tbw_4^*,F(\tbw_j)\tbw_j\rangle\tbw_4}{(\tE_j-\tE_4)\langle\tbw_4^*,\Lambda\tbw_4\rangle}.
\end{equation}

Proceeding to the next order of the expansions, at $\vep^3$ we obtain
\begin{eqnarray}
\label{W2}
\tE_j\Lambda\bW_j^{(2)} + e_j^{(2)}\Lambda\tbw_j &=& H\bW_j^{(2)}+F(\tbw_j)\bW_j^{(1)},
\end{eqnarray}
where we have used (\ref{W1}) and the property $F(\tbw_{3})=F(\tbw_{4})=0$ yielding
$F(\bW_j^{(1)})=0$.
Thus applying $\tbw_j^\dag$ to (\ref{W2}), we compute (recall $j=1,2$)
\begin{equation}
\label{W3}
e_j^{(2)}=\frac{1}{\langle\tbw_j^*,\Lambda\tbw_j\rangle}\sum_{p=3,4}\frac{\langle\tbw_p^*,F(\tbw_j)\tbw_j\rangle \langle\tbw_j^*,F(\tbw_j)\tbw_p\rangle}{(\tE_j-\tE_p)\langle\tbw_p^*,\Lambda\tbw_p\rangle}.
\end{equation}
Formulas (\ref{W1}) and (\ref{W3}) determine  the required terms of the expansions.

\subsection{Nonlinear modes bifurcating from $\tbw_{3,4}$}
\label{sec:bif34}
As it was mentioned above in our specific case $F(\tbw_{3,4})=0$ and thus the linear eigenvectors $\tbw_{3,4}$ at the same time solve the original nonlinear model (\ref{eq:stat}).
This means that the nonlinear modes (if any) bifurcating from $\tbw_{3,4}$  should in general possess finite nonzero   amplitude at the point of the bifurcation. In other words,   the bifurcations of nonlinear modes $\bw$   occur not from the linear limit (which is understood as $\|\bw\|\to0$), but rather from a finite amplitude solution (see also the discussion in~\cite{Mor1,Mor2}).
This readily
suggests that the small parameter expansions for nonlinear modes bifurcating from $\tbw_{3,4}$ should be looked for as follows ($j=3,4$):
\begin{eqnarray}
\label{expan_w34}
\bw_j  = \alpha_j\tbw_j + \vep  \bW_j^{(1)} + \vep^2\bW_j^{(2)}+\cdots,\\
\label{expan_E34}
E_j  =   \tE_j + \vep e_j^{(1)} +  \vep^2 e_j^{(2)}+\cdots,
\end{eqnarray}
where we have introduced the proportionality coefficients $\alpha_{j}$, which must be defined from   requirement of consistency of the asymptotic expansion. Notice that if $\alpha_j \ne 0$, then at the point of bifurcation (i.e. at $\vep=0$),  the power of the first   harmonic goes to zero, i.e. $|w_j^{(1,3)}| = 0$, while  $|w_j^{(2,4)}|$ stay finite.

Now one   has $F(\bw_j) = \vep F(\bW_j^{(1)}) + \vep^2 F(\bW_j^{(2)}) + \cdots$.
Substituting the introduced expansions into the nonlinear problem (\ref{eq:stat}), we observe that it is automatically satisfied in the leading order $\vep^0$.  In the $\vep^1$-order for
$j=3,4$ we obtain
\begin{equation}
\label{eq:vep1w34}
\alpha_j e_j^{(1)}\Lambda\tbw_j + \tE_j\Lambda\bW_j^{(1)} = H\bW_j^{(1)} + \alpha_j F(\bW_j^{(1)})\tbw_j.
\end{equation}
Employing again representation (\ref{exp_W}) [recall that $F(\bw_{3,4})=0$] we find $F(\bW_j^{(1)})=c_1 F(\tbw_1) + c_2 F(\tbw_2)$. Applying $\tbw_{7-j}^\dag$ to   both sides of (\ref{eq:vep1w34}) and using properties (\ref{ortt}) and
\begin{eqnarray}
\langle\tbw_{j}^*,F(\tbw_{1,2})\tbw_j\rangle=\langle\tbw_{7-j}^*,F(\tbw_{1,2})\tbw_j\rangle=0,
\end{eqnarray}
  we find that $c_{7-j}=0$. Next, we apply $\tbw_j^\dag$ and using the same arguments find that $e_j^{(1)}=0$. Finally, we apply $\tbw_{1}^\dag$  and $\tbw_{2}^\dag$ which yields the following system:
\begin{eqnarray}
c_{1} \langle\tbw_{1}^*,\Lambda \tbw_{1} \rangle (\tE_j -  \tE_{1})
 -\alpha_j c_1  \langle\tbw_{1}^*, F(\tbw_{1}) \tbw_j\rangle
 \nonumber \\%
 - \alpha c_2  \langle\tbw_{1}^*, F(\tbw_{2})  \tbw_j \rangle = 0,
 \\[2mm]
 c_{2} \langle\tbw_{2}^*,\Lambda \tbw_{2} \rangle (\tE_j -  \tE_{2})
  -\alpha_j c_1  \langle\tbw_{2}^*, F(\tbw_{1}) \tbw_j\rangle
  \nonumber \\%
  - \alpha_j c_2  \langle\tbw_{2}^*, F(\tbw_{2})  \tbw_j \rangle = 0.
\end{eqnarray}
For a given $j=3,4$ the latter equation form  a homogeneous linear system with respect to the coefficients $c_{1}$ and $c_{2}$. The compatibility condition of this system results in a \textit{quadratic} equation with respect to $\alpha_j$ whose roots give admissible values of $\alpha_j$ in  (\ref{expan_w34}). If $\alpha_j$ is chosen to satisfy the compatibility condition,  then  coefficients of $c_1$ and $c_2$ are given up to a multiplier, which however can be scaled out by means of renormalization of the small parameter $\vep$.

In order to determine $e_j^{(2)}$ one can proceed to the  $\vep^2$-order, which yields
\begin{eqnarray}
\alpha_j e_j^{(2)}\Lambda\tbw_j + \tE_j\Lambda \bW_j^{(2)} =
\hspace{3cm}
\nonumber \\%
 H\bW_j^{(2)}  + \alpha_j F(\bW_j^{(2)} ) \tbw_j + \alpha_j F(\bW_j^{(1)} )\bW_j^{(1)}.
\end{eqnarray}
After applying $\tbw_j^\dag$ one finds
\begin{equation}
 e_j^{(2)} = \frac{1}{\alpha_j \langle\tbw_{j}^*,\Lambda \tbw_{j} \rangle}\sum_{p=1}^2\sum_{q=1}^2 c_pc_q  \langle\tbw_{j}^*, F(\tbw_p)\tbw_{q} \rangle,
\end{equation}
which gives   the leading order   correction to the propagation constant.

\subsection{Discussion}

The results of two previous subsections show that   due to the interplay between the quadratic nonlinearity and the  particular structure of   eigenvectors of the  underlying linear problem, continuation of the linear eigenvectors into the nonlinear domain occurs  in two different ways, depending on whether the second or the first harmonic is vanishing in the linear eigenvector.  In the first case, when the total energy of the linear eigenvectors $\tbw_{1,2}$ is fully concentrated in the first harmonic, the bifurcations of nonlinear solutions from the linear ones resemble the standard Kerr nonlinearity case: each linear eigenstate gives birth to one family of nonlinear modes, and at the point of
the bifurcation ($\vep=0$) the amplitude of the nonlinear modes is zero, gradually increasing when one passes from $\vep=0$ to small nonzero $\vep$. Therefore, the bifurcations occur from the linear limit.

However, in the second situation, when the linear eigenvectors $\tbw_{3,4}$ have a
vanishing first harmonic contribution and a nonvanishing second harmonic
one, the bifurcations of nonlinear modes  occur in another way.  Now, at the point of bifurcation ($\vep=0$), the nonlinear modes generically bear a finite nonzero amplitude due to the presence of the additional term with the coefficient $\alpha_j$. Moreover, since the possible values of $\alpha_j$ are given by the quadratic equation, each linear eigenstate   in general gives birth  to at least two physically distinct families of nonlinear modes. Finally we notice that we have computed values of   $\alpha_j$ for several  choices of model parameters, obtaining in this way  analytical prediction for the   amplitude of   nonlinear modes at the point of   bifurcation ($\vep=0$).
For all cases that we checked  we   observed that  the analytically  predicted  amplitude of the bifurcating nonlinear modes agrees with that obtained from direct numerical results that follow in Sec.~\ref{sec:nonlin}.

\section{Fully Nonlinear Modes}
\label{sec:nonlin}
\subsection{Theoretical Setup}

Going beyond the consideration of the nonlinear modes  described by the
expansions (\ref{expan_w})--(\ref{expan_E}) and (\ref{expan_w34})--(\ref{expan_E34}), let us now turn to the set of all stationary  nonlinear modes  obeying the system (\ref{eq:stat}).
We observe that the latter   system has a considerable wealth of solutions of which we provide a representative set in what follows. In particular, we focus on nonlinear solutions
preserving the symmetry pertinent to the linear part, i.e. to the $\PT$-invariant modes obeying $\PT \bw = \bw$. Using for such modes  the amplitude-phase decomposition we rewrite the stationary solution $\bw$ introduced by (\ref{lin_ansatz}) in the form
\begin{eqnarray}
\bw=\left(
\begin{array}{c}
Ae^{i\phi_1}
\\
Be^{i\phi_2}
\\
Ae^{-i\phi_1}
\\
Be^{-i\phi_2}
\end{array}
\right),
\end{eqnarray}
where $A$ and $B$ are real stationary amplitudes and $\phi_{1,2}$
are stationary phases.
This ansatz reduces (\ref{eq:stat}) to the system of stationary equations as follows
\begin{subequations}
\begin{eqnarray}
       E& = &k_1e^{-2i\phi_1} - {2}Be^{i(\phi_2 -2\phi_1)} + i\gamma_1,\\
    {2}E& = &k_2e^{2i\phi_2} - (A^2/B) e^{i(\phi_2 -2\phi_1)} - q  - i\gamma_2
\end{eqnarray}
\end{subequations}
(where it is assumed that $B\neq 0$). Further splitting to real and imaginary parts yields four equations:
\begin{subequations}
\label{eq:sys4}
\begin{eqnarray}\label{nmsys}
E &=& k_1\cos(2\phi_1) -  {2}B\cos(\phi_2 - 2\phi_1),\\
0 &=& -k_1\sin(2\phi_1) -  {2}B\sin(\phi_2 - 2\phi_1) + \gamma_1,\\
 {2}E &=& k_2\cos(2\phi_2) - (A^2/B)\cos(\phi_2 - 2\phi_1)-q,\\
0 &=& k_2\sin(2\phi_2) - (A^2/B)\sin(\phi_2 - 2\phi_1) - \gamma_2.
\end{eqnarray}
\end{subequations}

If we take $A$, $B$, $\phi_{1,2}$ as four unknowns in the system
(\ref{eq:sys4}), then  one can expect that there exists one or
several solutions for any given $E$. Therefore, we can speak about
{\em continuous families} of nonlinear modes.
In order to visualize these families, one can
introduce the quantity
\begin{eqnarray}
\label{energy}
U=\langle \bw,\Lambda\bw \rangle=\bw^\dag\Lambda\bw
\end{eqnarray}
which corresponds to the Manley-Rowe invariant (of the conservative system $\gamma_{1,2}=0$ where it is a conserved quantity). In the case at hand we have
$U= 2(A^2+ 2 B^2)$.
Then the families of the nonlinear modes can be displayed as
dependencies on the $(E,U)$ plane  as the functions $U$ \textit{vs} $E$.

\subsection{Numerical results}
\begin{figure}
{\includegraphics[width=1.0\columnwidth]{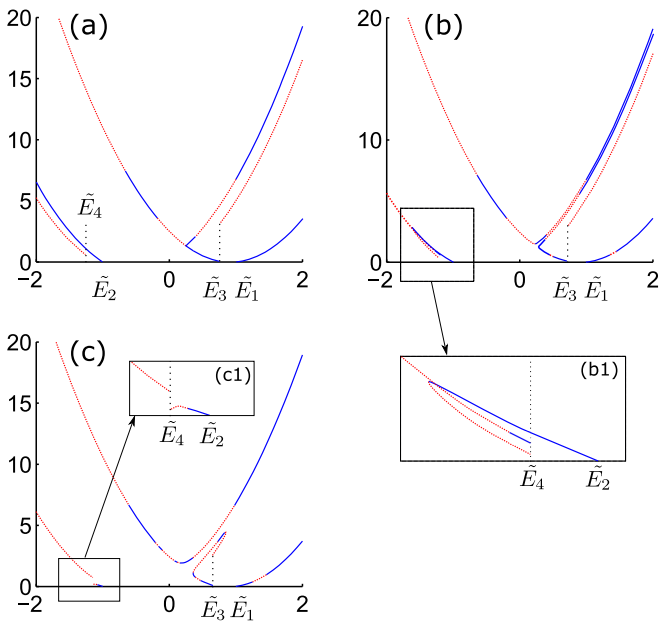}}%
\caption{(Color online)  Families of
nonlinear modes on the plane $U$   \textit{vs} $E$.
Values of parameters are given as
  $k_1=1$, $k_2 =2$, $q =
0.5$, and  $\gamma_{1,2}=0$ [panel (a)];  $\gamma_1=0.1$ and $\gamma_2
= 0.5$  [panel (b)];  $\gamma_1=0.1$ and $\gamma_2=0.9$  [panel (c)]. Stable and  modes correspond to the blue (solid)  and  red (dashed) fragments of the curves. Insets (b1) and (c1) provide better resolution for some bifurcational features   not visible well in the main plots.}
\label{fams}
\end{figure}

Several examples  are presented in Fig.~\ref{fams} where  we first
address the situation $\gamma_{1,2}=0$ corresponding to the
conservative limit of the problem and then consider the effect of
nonzero $\PT$-symmetric components $\gamma_{1,2}$ for some
representative value pairs.

The nonlinear system (\ref{eq:sys4}) features several interesting properties. First, as it was predicted above
there exist families bifurcating from the linear eigenstates $\tbw_{1,2}$  [see
Eqs.~(\ref{E}) and (\ref{eq:w12})], one family bifurcating from each eigenstate.  In accordance with the expansions (\ref{expan_w})--(\ref{expan_E}), at the points of the bifurcations ($\vep=0$) one has $E=\tE_{1,2}$ and $U=0$. The modes obeying expansions (\ref{expan_w34})--(\ref{expan_E34}) have also been found in our numerics. For such modes at $\vep=0$ one has  $E=\tE_{3,4}$ and $U=4\alpha_{3,4}^2$   with $\alpha_{3,4}$ being solutions of the quadratic equation introduced in Sec.~\ref{sec:bif34} [in order to obtain the latter equality we used Eqs.~(\ref{eq:w34}) for the explicit form of  eigenvectors $\tbw_{3,4}$]. In all the considered cases  we have found two distinct families bifurcating either from $\tE_3$ or from $\tE_4$.
{Notice however, that the two families bifurcating from $\tE_4$ for $\gamma_{1,2}=0$ are not distinguishable in Fig.~1(a). This is because for each given $E$ the modes belonging to  those families are mutually complex conjugate. Therefore, these solutions have the same $U$ characteristic. They also have the identical stability properties (see below for the discussion on stability) which allows us to consider only one family of those  two in what follows.}

It is also important to indicate that in all cases addressed in Fig.~\ref{fams}, the values of
the parameter $\alpha_{3,4}$ allowing for the bifurcations from $\tE_{3,4}$ are distinct from zero.   Therefore,  the  value  $U=4\alpha_{3,4}^2$ corresponding to the point of bifurcation   is also distinct from zero.  The latter comment is relevant because in all   three panels of Fig.~\ref{fams} one can observe that one of the  emerging at $E=\tE_3$ families approaches closely the horizontal axis $U = 0$. However, as the above analysis clearly indicates, the relevant bifurcation point is still distinct from $U = 0$.

In terms of gross
features of the bifurcation diagrams, we observe that the families
are extended to the domains of either
positive or negative $E$, and some families feature a
parabolic-shaped pattern. We also notice
that an apparent
pitchfork bifurcation existing in   panel (a) with
$\gamma_{1,2} =0$ breaks into a pair of ``fold
points'', as we deviate from the Hamiltonian limit. Additional
such fold points can be observed e.g. in   panel (c) of
Fig.~\ref{fams} for $E\approx 0.85$, or in the   panel (b)
 for $E\approx -1.63$ [see inset (b1)]. However, it is evident that most families in the figure either come from $-\infty$ or asymptote towards $+ \infty$,
for large values of $U$.

In Fig.~\ref{fams} we also address stability of the obtained modes
by means of identifying of the spectrum of the
linearization of the original equations around each family of
stationary modes. One can observe that the stability situation may be fairly
complicated with the same family  having alternating domains of
stability and instability (these will be analyzed in detail below
for the parameters used).
The linearization spectrum contains a
double zero eigenvalue (due to the global phase invariance of the full model).
Instability can be caused either by a pair
of purely real eigenvalues in the spectrum (one of them is
responsible for instability) or by a quartet of complex
eigenvalues (two of which correspond to unstable modes).

\begin{figure}
 {\includegraphics[width=1.\columnwidth]{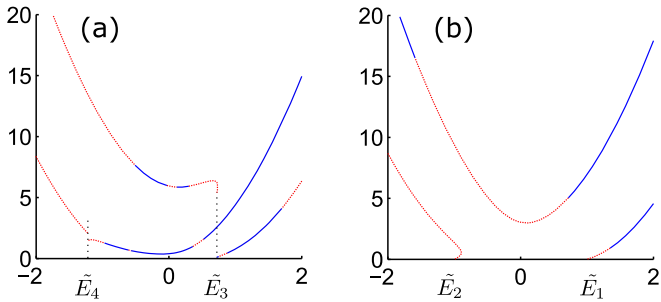}}%
 \caption{(Color online)  Families of
nonlinear modes on the planes $U$   \textit{vs} $E$   when the $\PT$ symmetry of the underlying linear
problem is broken.  Panel (a): regime (ii) with
$\gamma_1 = 1.1$ and $\gamma_2 =
0.5$;  panel (b): regime (iii) with
$\gamma_1 = 0.1$, $\gamma_2 = 2.1$ [see Sec.~\ref{linear} for definition of regimes (ii) and (iii)]. All other
parameters are the same as in Fig.~\ref{fams}. Notice that eigenvalues $\tE_{1,2}$ are complex in panel (a), while $\tE_{3,4}$ are complex in panel (b).}
\label{fams-broken}
\end{figure}

Quite remarkably, nonlinear modes (including stable ones) can be
also found in the regime where
the $\PT$ symmetry of the underlying linear problem
is broken. This is a feature that nonlinearity has been
shown to sustain even in the cubic case, in particular for a quadrimer setting (see e.g. the relevant discussion of~\cite{pgk,konorecent3}).
Two examples are shown in Fig.~\ref{fams-broken}. The
respective parameters correspond to the underlying linear problem
belonging to two different ``phases'' of the broken $\PT$ symmetry
(see Sec.~\ref{linear}). Namely, the system belongs to the phase
(ii) for  panel (a) while   panel (b) corresponds to the
phase (iii).  Respectively,  in the situation of panel (a) the eigenvalues $\tE_{1,2}$  are complex (with nonzero imaginary parts) and do not allow for bifurcations of nonlinear modes. However, the eigenvalues $\tE_{3,4}$ are still real and give birth to the families of solutions (two families emerge at  $\tE_3$ and
at $\tE_4$). \textit{Vice versa},  in panel (b) the eigenvalues $\tE_{3,4}$ are complex but one can observe families bifurcating from the  real eigenvalues $\tE_{1,2}$ (one family from each eigenvalue).
Solutions arising from $\tE_{1,2}$ are unstable in the vicinity of the bifurcations, but  sufficiently strong
nonlinearity in this case is critical for enabling dynamical stability.
Practically, it also appears that the modes where the Manley-Rowe
invariant has a positive slope have a wider stability interval, although
a more quantitative observation along these lines is, presently, absent.

\begin{figure}
\scalebox{0.28}{\includegraphics{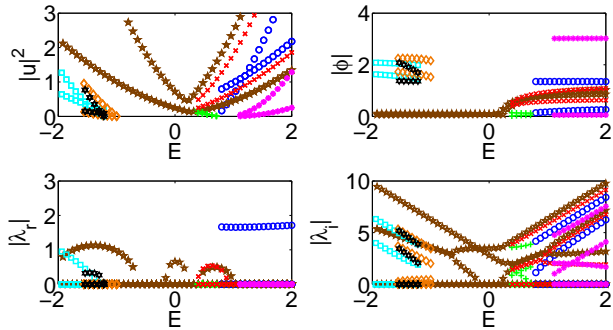}}
\caption{(Color online) Existence and stability properties of nonlinear modes with $k_1=1$, $k_2 =2$, $q = 0.5$, $\gamma_1=0.1,\ \gamma_2 = 0.5$.
The four panels denote the solution amplitude
(top left), phase differences between adjacent nodes
(top right),  real and imaginary parts (second row) of eigenvalues.
For a detailed explanation of the different families, see the text.}
\label{chisq_k12q5g15}
\end{figure}

\begin{figure}[htp]
\scalebox{0.27}{\includegraphics{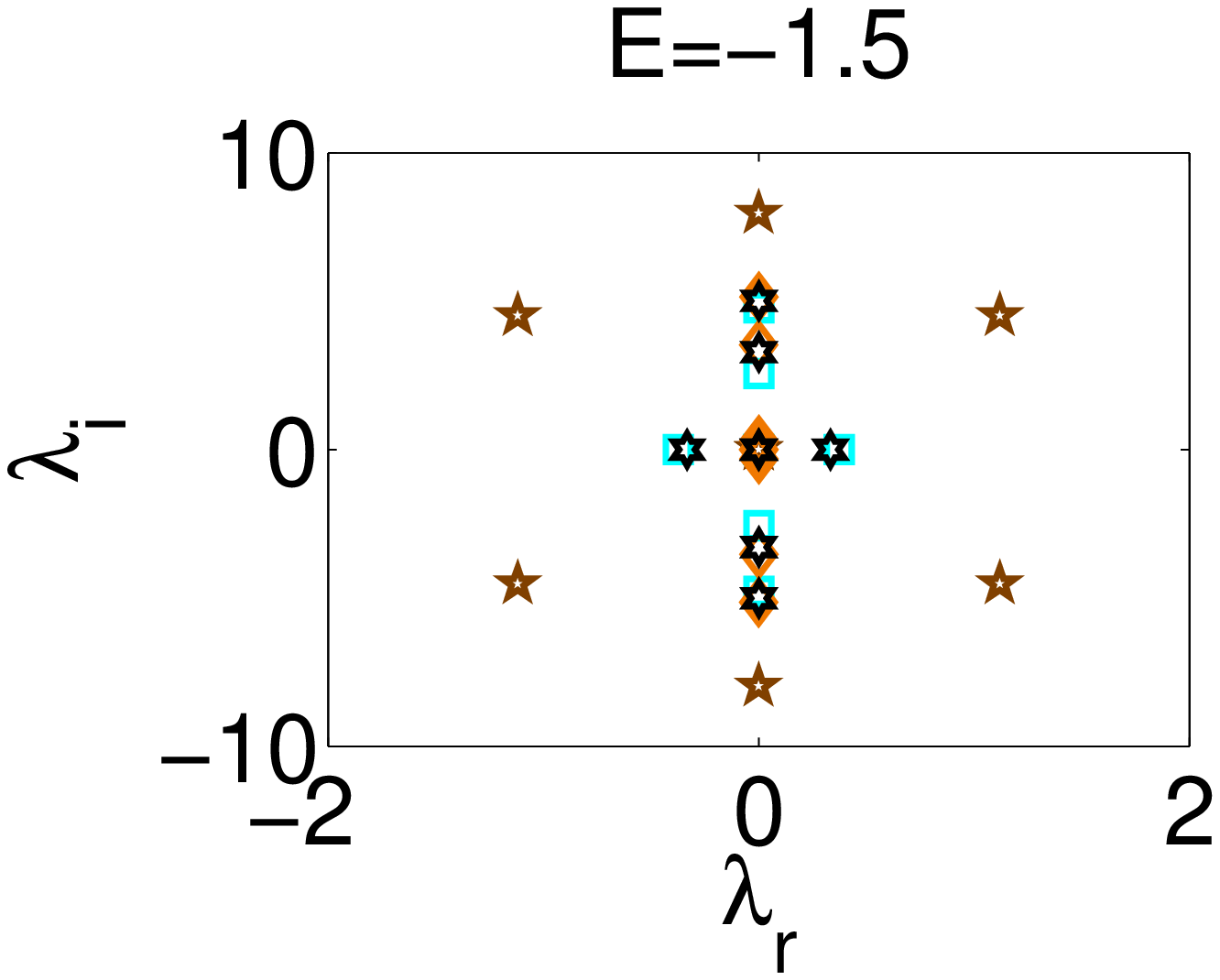}}
\scalebox{0.27}{\includegraphics{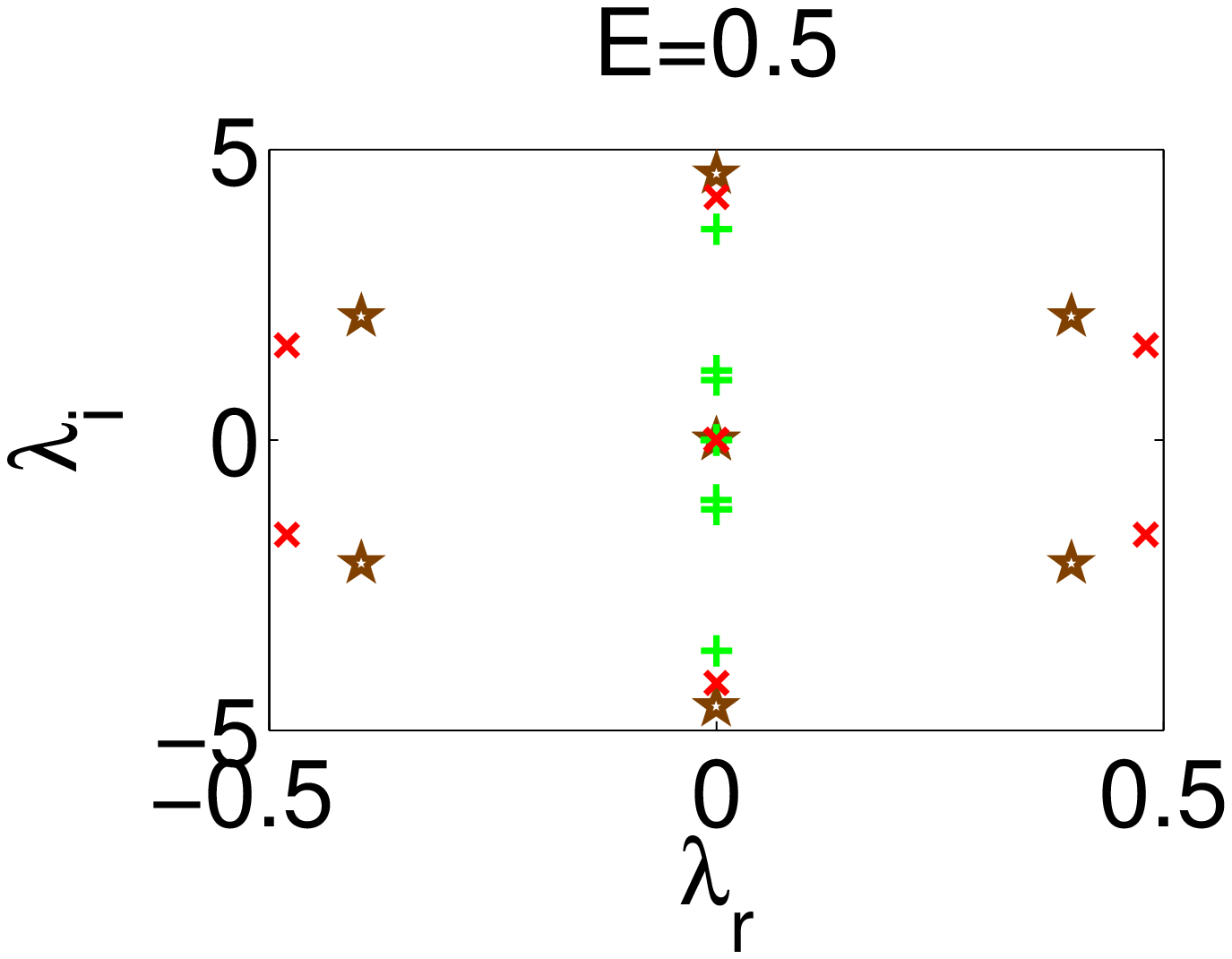}}
\scalebox{0.27}{\includegraphics{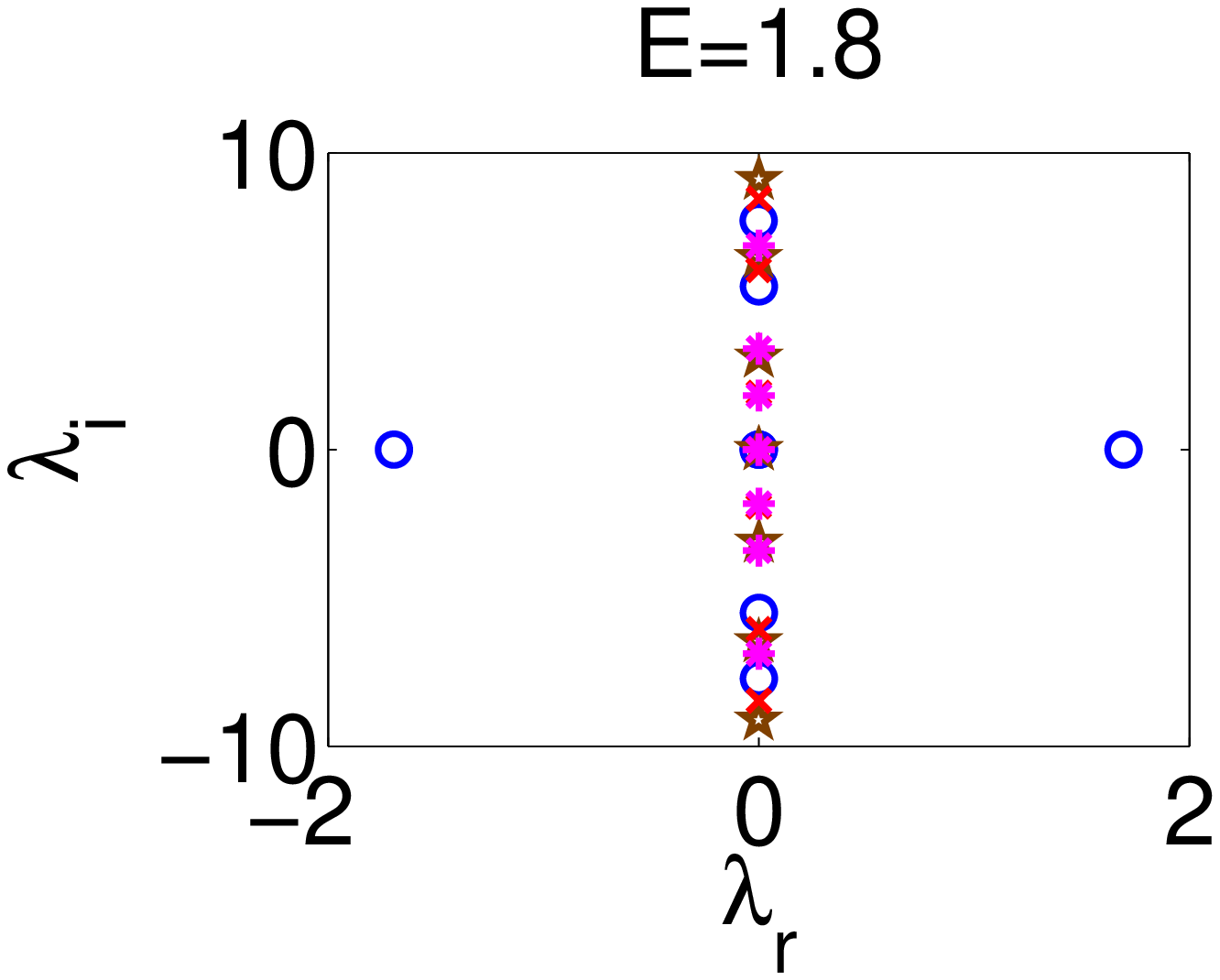}}
\caption{(Color online) Eigenvalues of the linearization problem of nonlinear modes with $k_1=1$, $k_2 =2$, $q = 0.5$, $\gamma_1=0.1,\ \gamma_2 = 0.5$. The same notation
has been used as in Fig.~\ref{chisq_k12q5g15}.}
\label{chisq_k12q5g15_stab}
\end{figure}

We now turn to a more systematic analysis of the existence and stability
properties of the different  families of solutions identified previously, for
reasons of completeness.
Fig.~\ref{chisq_k12q5g15} illustrates the situation where $k_1=1$, $k_2 =2$, $q = 0.5$, $\gamma_1=0.1,\ \gamma_2 = 0.5$.
There are eight families of solution in this case, denoted by different symbols.
Their eigenvalues for the respective parameters of
existence are shown in the case of three different choices
of $E$ in Fig.~\ref{chisq_k12q5g15_stab}.

\begin{itemize}
\item The family denoted by blue circles arises from $E=\tE_3 \approx 0.72$ and continues monotonically  increasing
its Manley-Rowe diagnostic upon increase of $E$ to infinity.
It always has two pairs of purely imaginary and
 one pair of real eigenvalues, which give rise to its instability.

\item The brown pentagram family exists for all the considered
values of $E$. The amplitudes of both harmonics reach their minimum
(within the parabolic shape of the family reported previously) in
the interval $E \in [0.2,0.3]$, but at different
 points. This family has three pairs of purely imaginary eigenvalues, two of which collide at $E\approx -2.13$ and turn into a complex quartet.
 At $E \approx -0.65$, the complex quartet collides on the imaginary axis and
splits anew into two pairs of imaginary eigenvalues, restabilizing the
waveform. The larger  of the two imaginary
 pairs subsequently meets the largest imaginary eigenvalues
and the collision yields a complex quartet within the short
parametric interval of $E \in [-0.20,-0.19]$ (hereafter boundaries of the intervals are given approximately). The remaining (lowest frequency)
pair  collides with the spectral plane origin and turns into a real pair
at $E\approx -0.2$. This pair of eigenvalues becomes imaginary again shortly at
$E\approx 0.2$ and collides
  with its former partner at $E\approx 0.41$ to form a complex quartet. This complex quartet once again splits into two purely imaginary pairs
  at $E\approx 1$. As a result, the brown pentagrams family is stable for all $E$ except on $[-2.13,-0.65], [-0.2,0.2], [0.41,1]$.
From the above, the substantial complexity of the family stability properties
should be rather evident.

\item The green pluses and the red crosses arise together from a saddle-node bifurcation at $E\approx 0.3$. The green pluses family is essentially
 stable except when $E$ is within a small interval of $[0.47,0.48]$,
where two out of three pairs of purely imaginary eigenvalues collide
yielding a Hamiltonian-Hopf bifurcation and a
 complex quartet and the reverse path renders the eigenvalues
purely imaginary again. This family terminates at $E=\tE_3 \approx 0.72$  with  the first harmonic amplitude vanishing.

\item The red crosses family bifurcates from the same point as the green pluses,
however it does not terminate. It is   unstable only on an interval
of $E \in [0.39,0.89]$ due to a complex quartet.

\item The magenta stars family arises from the linear limit at $E = \tE_1 \approx 0.99$ and
exists always thereafter. It has three pairs of purely imaginary eigenvalues, too. Two of them turn into a complex quartet within the
small interval $[1.38,1.43]$ and make the family unstable in this interval.

\item The cyan squares family comes from $-\infty$ having a real pair and two purely imaginary pairs of eigenvalues. This branch is stable only after $E\approx -1.3$ where the real pair turns purely imaginary; subsequently the branch terminates at $E=\tE_4 \approx -1.22$
(with  the first harmonic amplitude vanishing).

\item The orange diamonds and the black hexagrams   emerge from a saddle-node bifurcation at $E\approx -1.63$. The orange diamonds constitute
 the only family that is always stable, having three pairs of purely
imaginary eigenvalues. This family terminates at the linear limit of
$E=\tE_2 \approx -0.99$.

\item The black hexagrams start at the same point as the orange diamonds but terminates at $E=\tE_4 \approx -1.22$. It always has two pairs of purely imaginary and one pair of real eigenvalues. Hence it is generically unstable.
\end{itemize}
As general comments we can infer that, arguably, the most robust families
and ones that will generically exist are the ones emerging from the
eigenvalues  $\tE_{1,2}$ of the linear limit. The other families may have intervals of
stability but also often suffer oscillatory or real instabilities and
are subject to saddle-center bifurcations (although e.g., the family
starting from $\tE_1 \approx 0.99$ also has a small interval of instability, and  the generically stable family
starting from $\tE_2 \approx -0.99$ is subject to a saddle-node bifurcation).

\begin{figure}
\scalebox{0.28}{\includegraphics{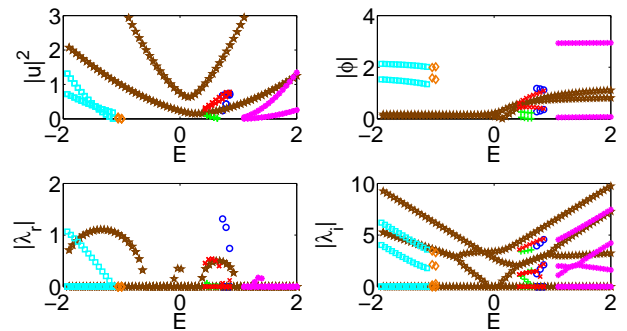}}
\caption{(Color online)  Existence and stability properties of nonlinear modes with similar settings as in Fig.~\ref{chisq_k12q5g15} but for $k_1=1$, $k_2 =2$, $q = 0.5$, $\gamma_1=0.1,\ \gamma_2 = 0.9$.}
\label{chisq_k12q5g19}
\end{figure}

\begin{figure}
\scalebox{0.27}{\includegraphics{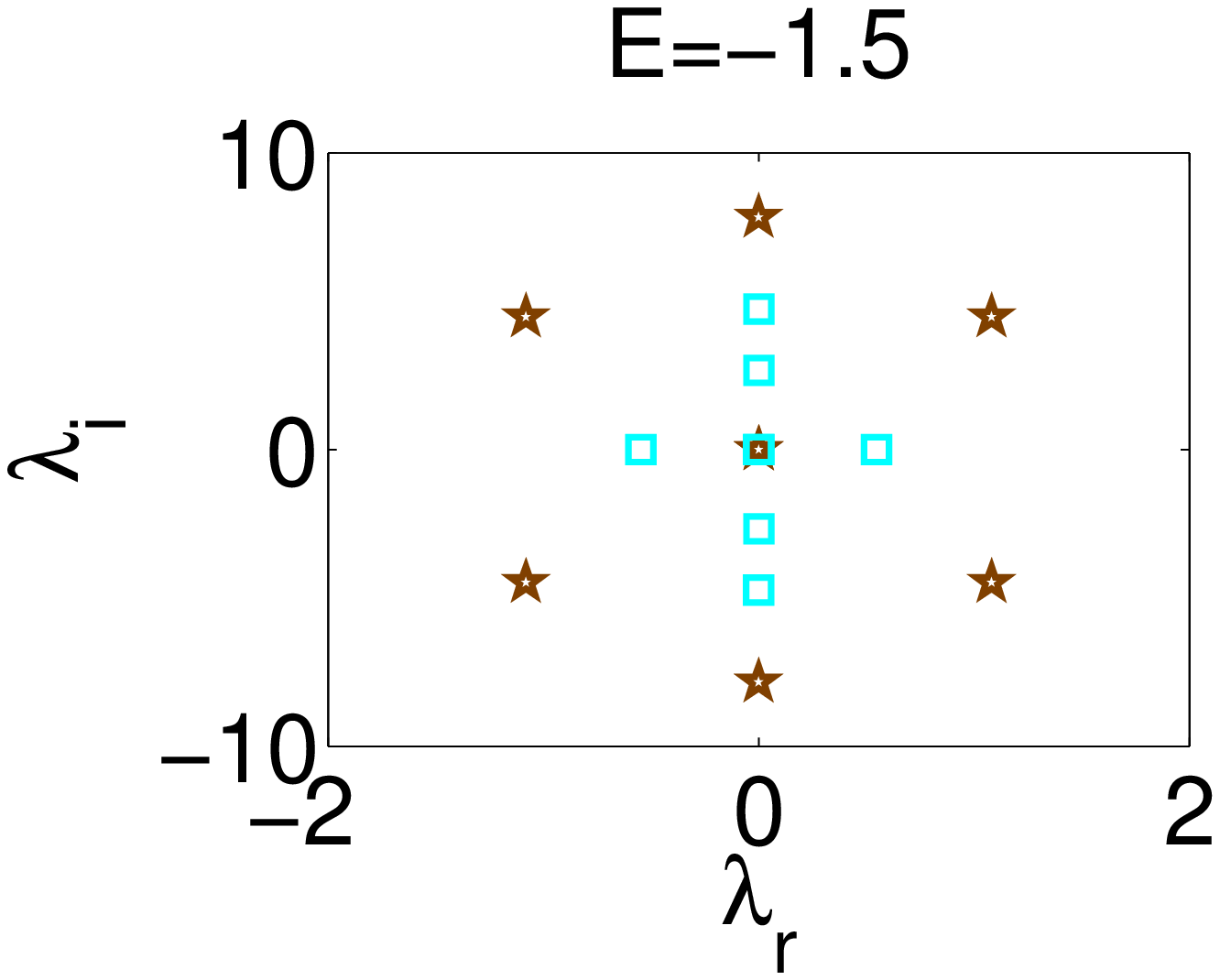}}
\scalebox{0.27}{\includegraphics{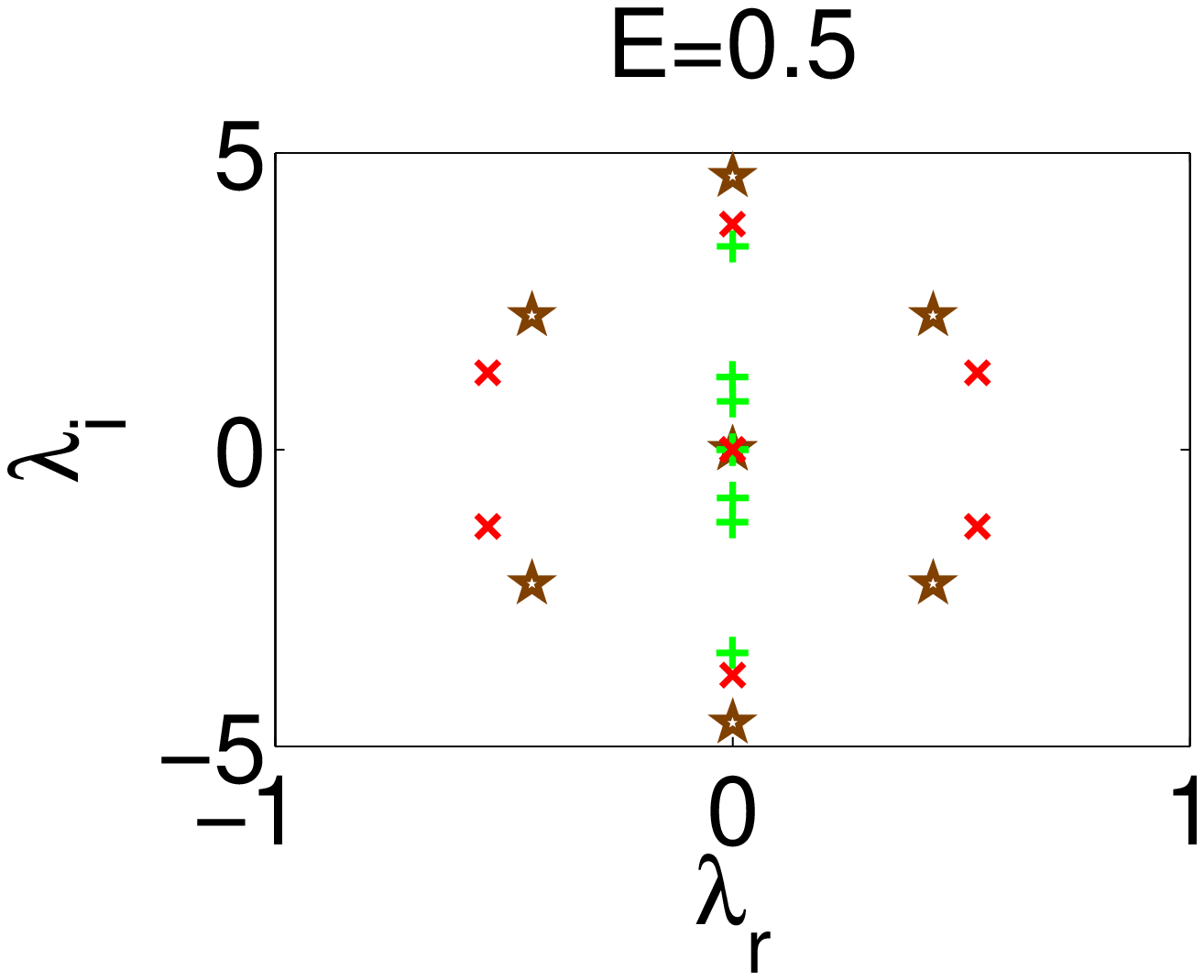}}
\scalebox{0.27}{\includegraphics{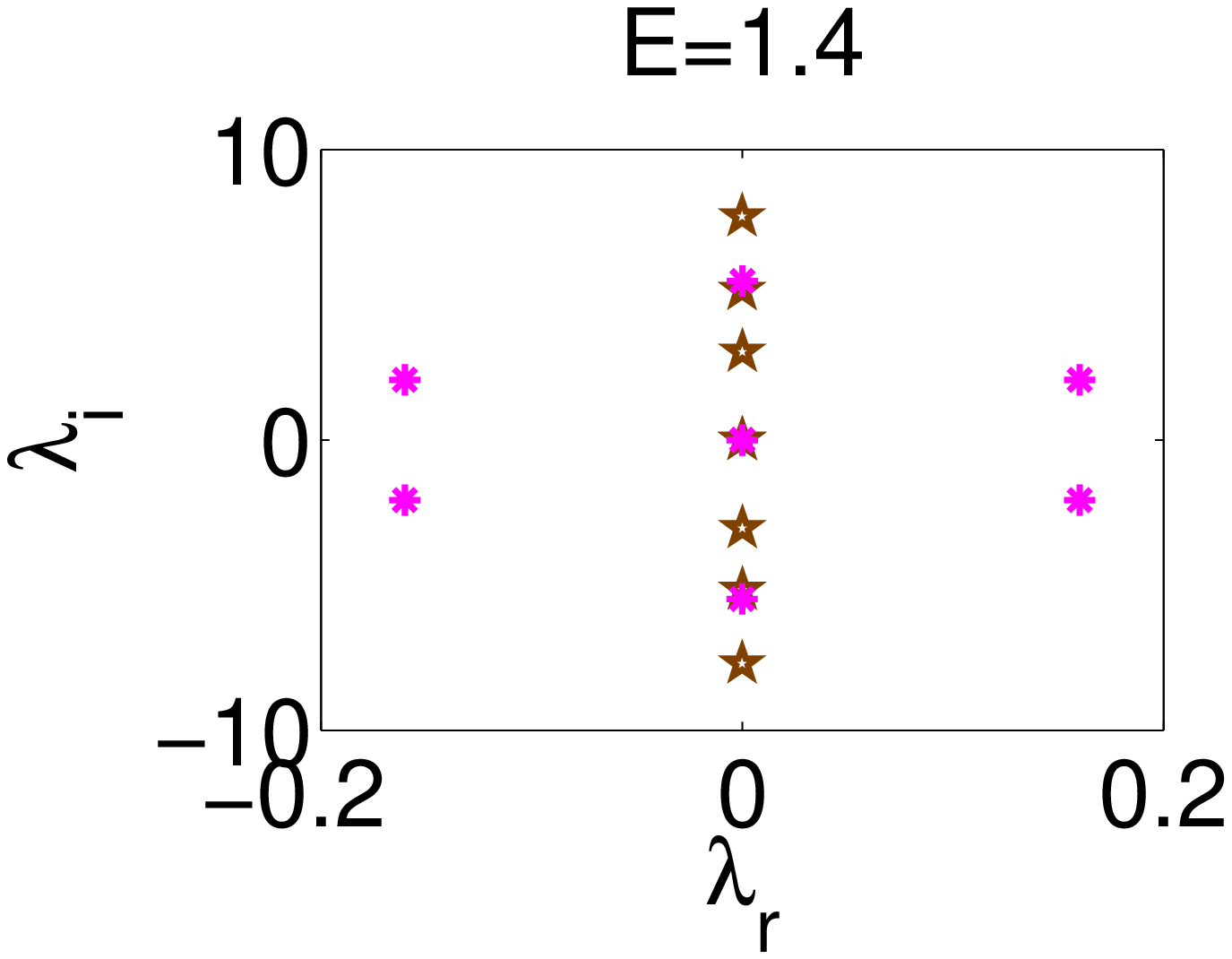}}
\caption{(Color online)  Eigenvalues of the linearization problem of nonlinear modes with $k_1=1$, $k_2 =2$, $q = 0.5$, $\gamma_1=0.1,\ \gamma_2 = 0.9$.}
\label{chisq_k12q5g19_stab}
\end{figure}

Fig.~\ref{chisq_k12q5g19} and Fig.~\ref{chisq_k12q5g19_stab} show us the solution profiles and their eigenvalues under the parameter $k_1=1$, $k_2 =2$, $q = 0.5$, $\gamma_1=0.1,\ \gamma_2 = 0.9$. In this case there are seven families. The black hexagrams family of Fig.~\ref{chisq_k12q5g15} does not exist any more. We briefly summarize the difference in each family in the following compared with the previous ones.
\begin{itemize}
\item The red crosses family now arises from a saddle-node bifurcation with the green pluses at $E\approx 0.36$ and terminates into another saddle-node bifurcation with the blue circles families at $E \approx 0.85$. It now has a pair of purely imaginary and a complex quartet eigenvalues. The latter one reshapes into two pairs of purely imaginary eigenvalues at $E\approx 0.74$, and one of
them becomes real at
$E\approx 0.84$. Hence, it is unstable except on the interval $[0.74,0.84]$.

\item The blue circle branch is still unstable but now exists from $E=\tE_3 \approx 0.64$ to $E\approx 0.85$.

\item The green pluses branch
now exists from $E\approx 0.36$ to $\tE_3 \approx 0.64$. It is essentially stable except when $E$ is between $[0.4,0.46]$.

\item The brown pentagrams still exist for all $E$ and bear similar eigenvalues as in Fig.~\ref{chisq_k12q5g15}. In this case, the
branch is stable except on $[-2.06,-0.61],[-0.22,-0.17],[-0.1,0.07],[0.37,0.97]$.

\item The magenta stars family is similar as in Fig.~\ref{chisq_k12q5g15},
again bifurcating from the linear limit and
now being stable in the exception of the interval $E \in [1.27,1.45]$.

\item The unstable cyan squares family still comes from $-\infty$, but now it is always unstable and terminates at $\tE_4=-1.14$.

\item The orange diamonds family exists from $\tE_4\approx -1.14$ to $\tE_2 \approx -0.99$. It is unstable until $E\approx -1.1$ and becomes stable thereafter.
\end{itemize}

\begin{figure}
\scalebox{0.28}{\includegraphics{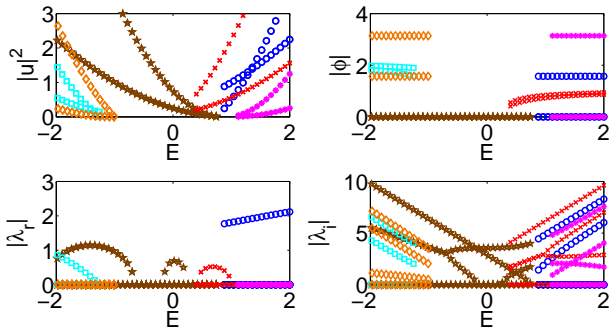}}
\caption{(Color online)  Existence and stability properties
of nonlinear modes with similar settings as Fig.~\ref{chisq_k12q5g15} but for $k_1=1$, $k_2 =2$, $q = 0.5$, $\gamma_1=0,\ \gamma_2 = 0$.}
\label{chisq_k12q5g00}
\end{figure}

\begin{figure}
\scalebox{0.27}{\includegraphics{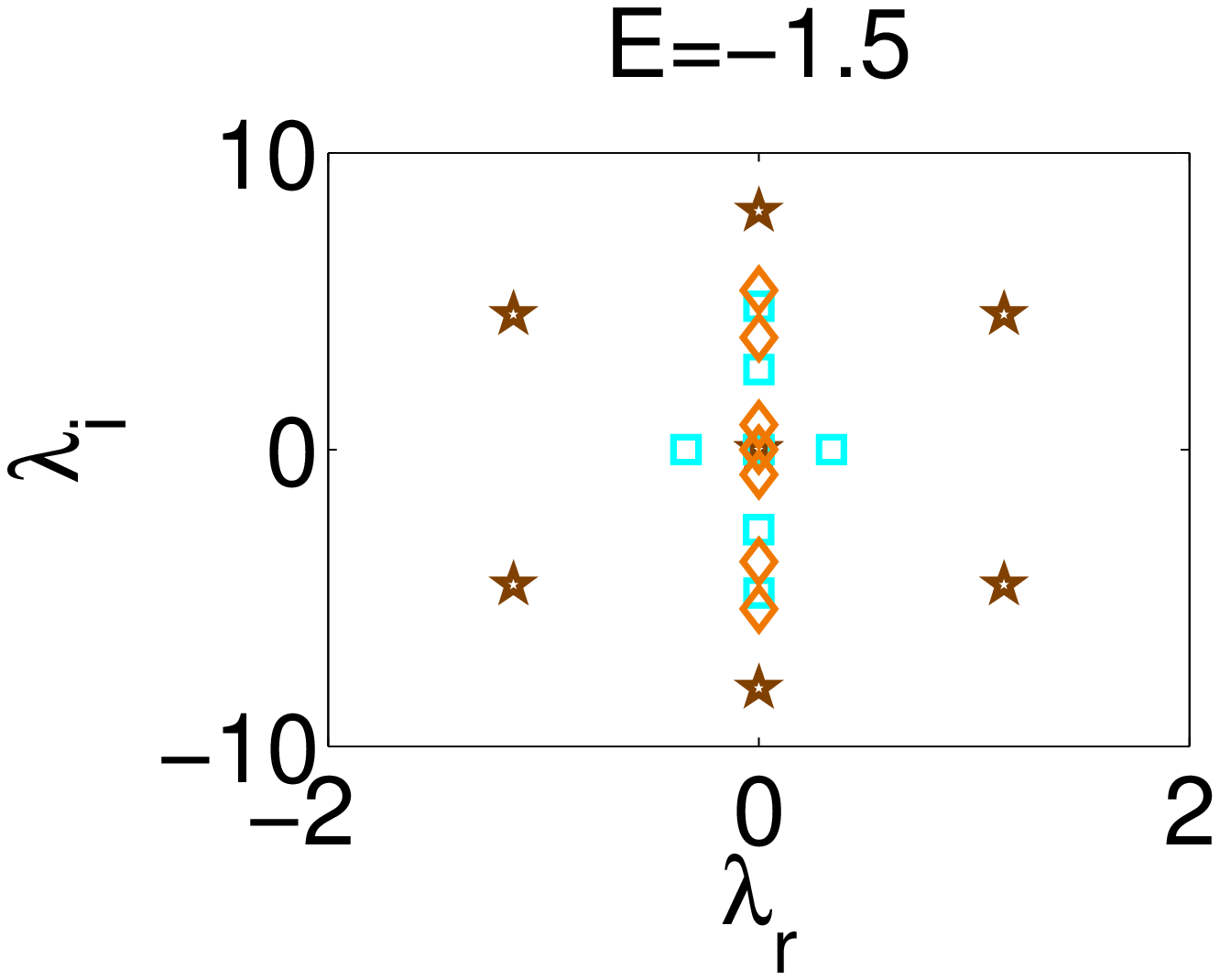}}
\scalebox{0.27}{\includegraphics{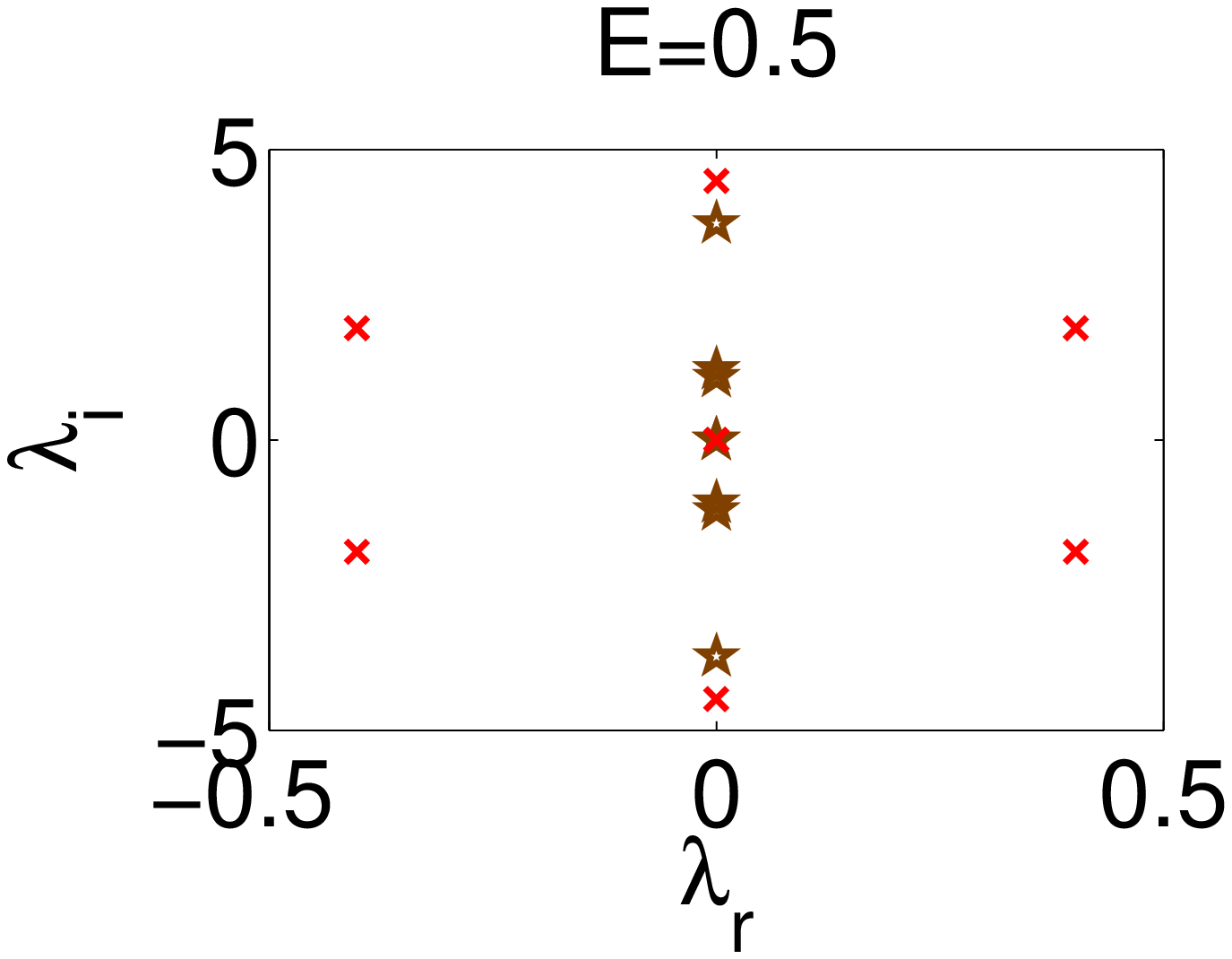}}
\scalebox{0.27}{\includegraphics{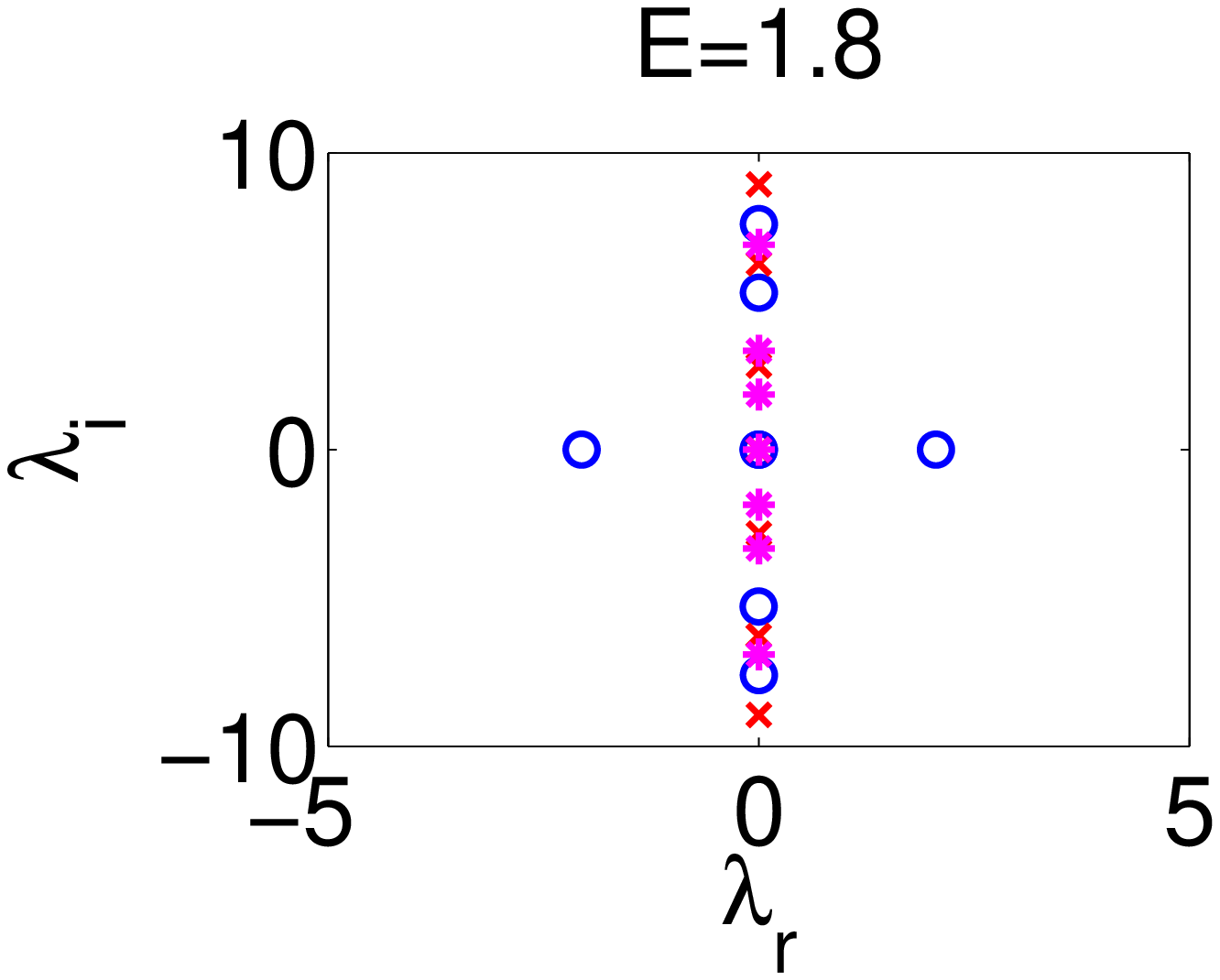}}
\caption{(Color online) Eigenvalues of the linearization problem of nonlinear modes with $k_1=1$, $k_2 =2$, $q = 0.5$, $\gamma_1=0,\ \gamma_2 = 0$.}
\label{chisq_k12q5g00_stab}
\end{figure}

For comparison purposes, we also consider the
Hamiltonian case  $k_1=1$, $k_2 =2$, $q = 0.5$, $\gamma_1=0,\ \gamma_2 = 0$.
Fig.~\ref{chisq_k12q5g00} and Fig.~\ref{chisq_k12q5g00_stab} show the six families of nonlinear modes in this case.
\begin{itemize}
\item The blue circle family is similar to the one in Fig.~\ref{chisq_k12q5g15}, i.e.   arises from $E=\tE_3 = 0.75$
and is always unstable. It possesses a pair of real and two pairs of purely imaginary eigenvalues for all $E$ where it exists.

\item The brown pentagrams now exist only up to $\tE_3=0.75$.
This branch is stable except on $[-2.17,-0.66], [-0.17,0.24]$, where it has a
complex quartet of eigenvalues.

\item The red crosses now bifurcate from the brown pentagrams at $E\approx 0.25$ and persist beyond the point.
It is this bifurcation that apparently splits into two fold points in
the two cases considered previously.
The red crosses are unstable only on the interval $[0.41,1]$,
where a complex quartet of eigenvalues comes
from two pairs of purely imaginary ones
colliding at $E\approx 0.41$ and returning to the imaginary axis at $E\approx 1$.

\item The magenta stars family still arises from the linear limit at $\tE_1=1$. However, it now always has three pairs of purely imaginary eigenvalues and hence is
stable wherever it exists.

\item The cyan squares family is similar to the one in Fig.~\ref{chisq_k12q5g15}, too. It is unstable, comes from $-\infty$, and terminates at $\tE_4=-1.25$.

\item The orange diamonds branch now also exists from $-\infty$ and terminates at $\tE_2=-1$. It is always stable in this case, too, again verifying the robustness
of the families that emerge from the linear limit.
\end{itemize}

\section{Dynamics of the system}
\label{sec:dyn}
Finally, from the point of view of numerical results,
we have also performed   direct numerical simulations of the propagation
dynamics of the quadratically nonlinear $\PT$-symmetric dimer.
These simulations allow us to obtain a feeling about the
dynamical implications of the instabilities presented above.

\begin{figure}
\subfigure[\ blue circles]{\scalebox{0.28}{\includegraphics{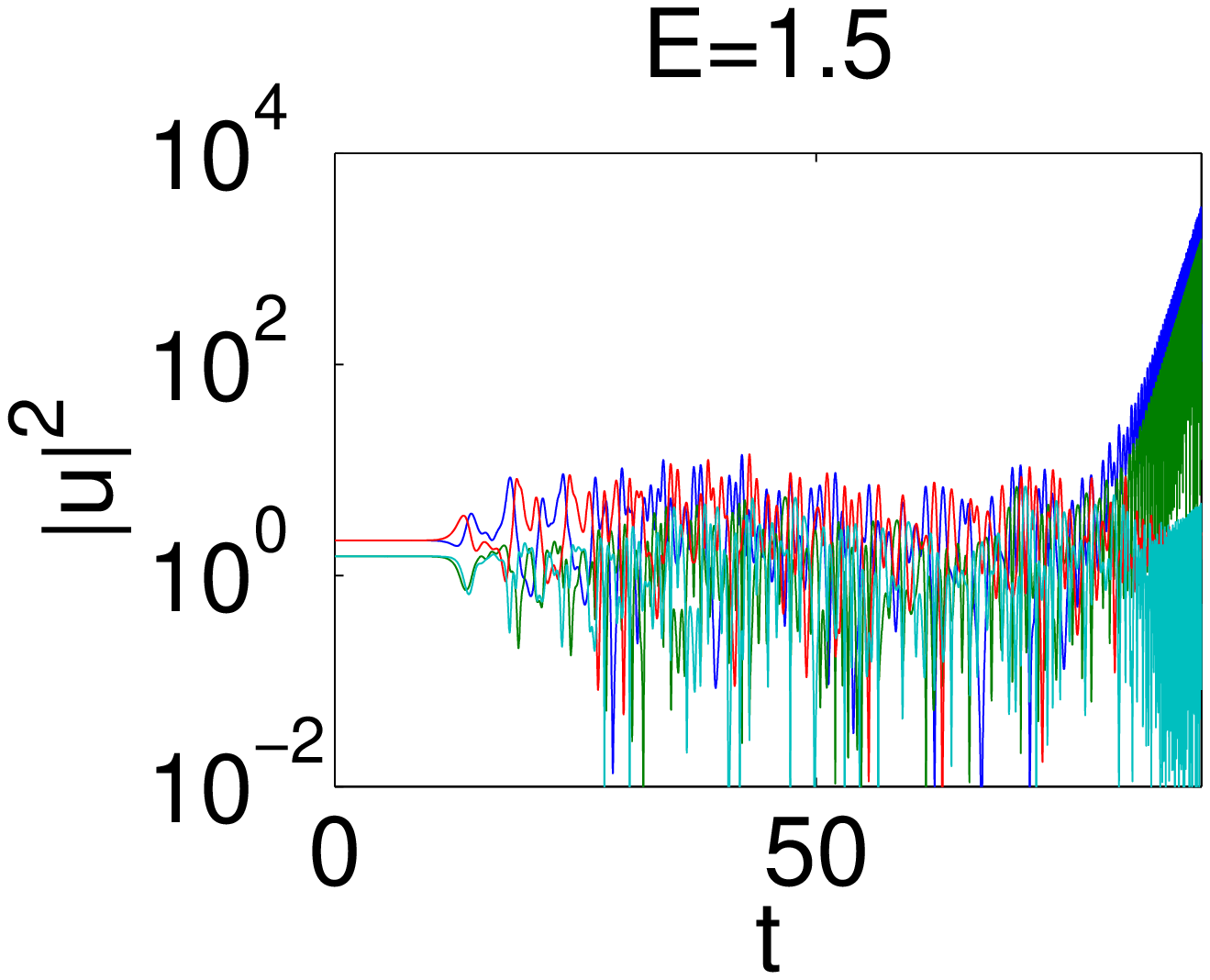}}}%
\subfigure[\ brown pentagrams]{\scalebox{0.28}{\includegraphics{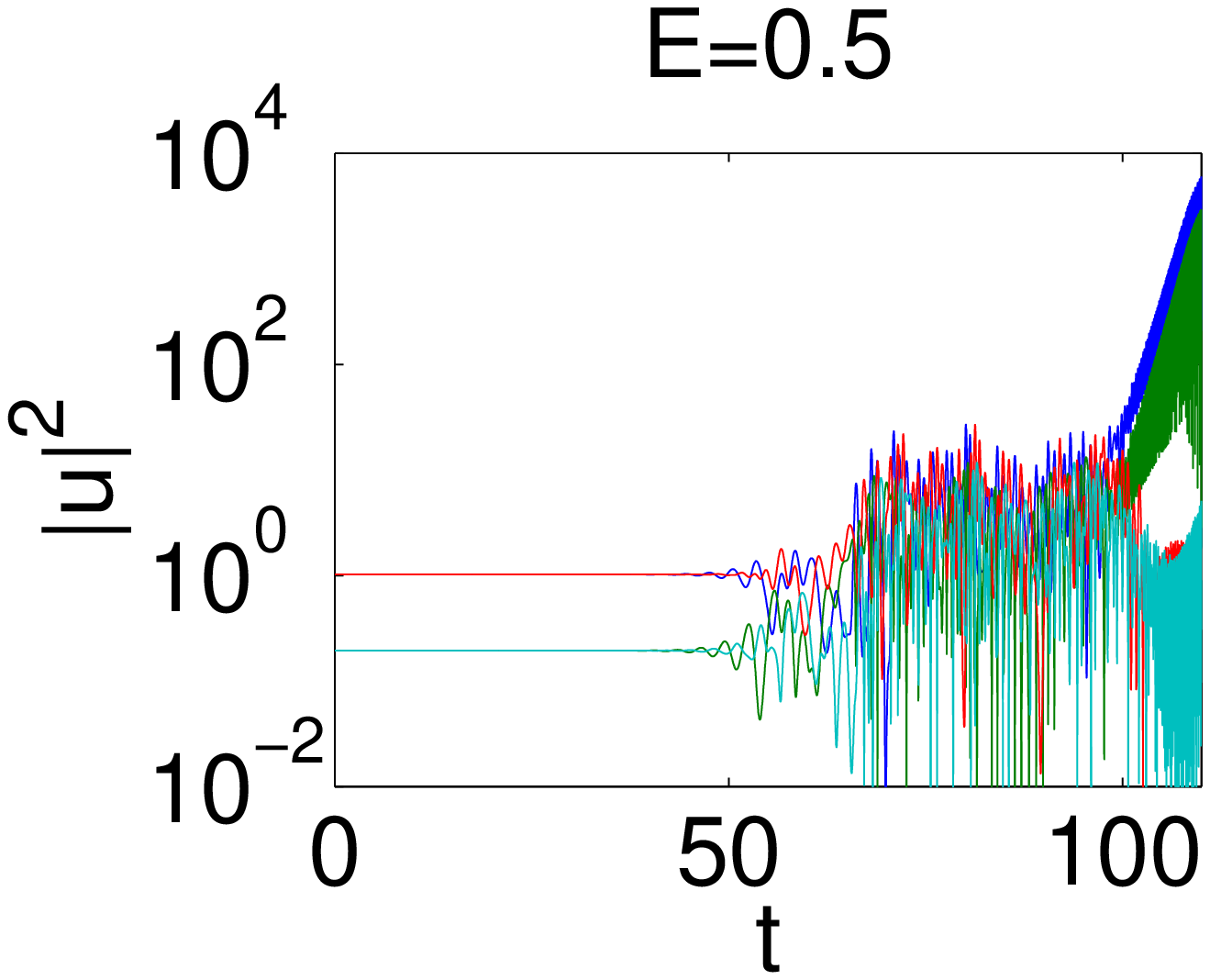}}}\\
\subfigure[\ green pluses]{\scalebox{0.28}{\includegraphics{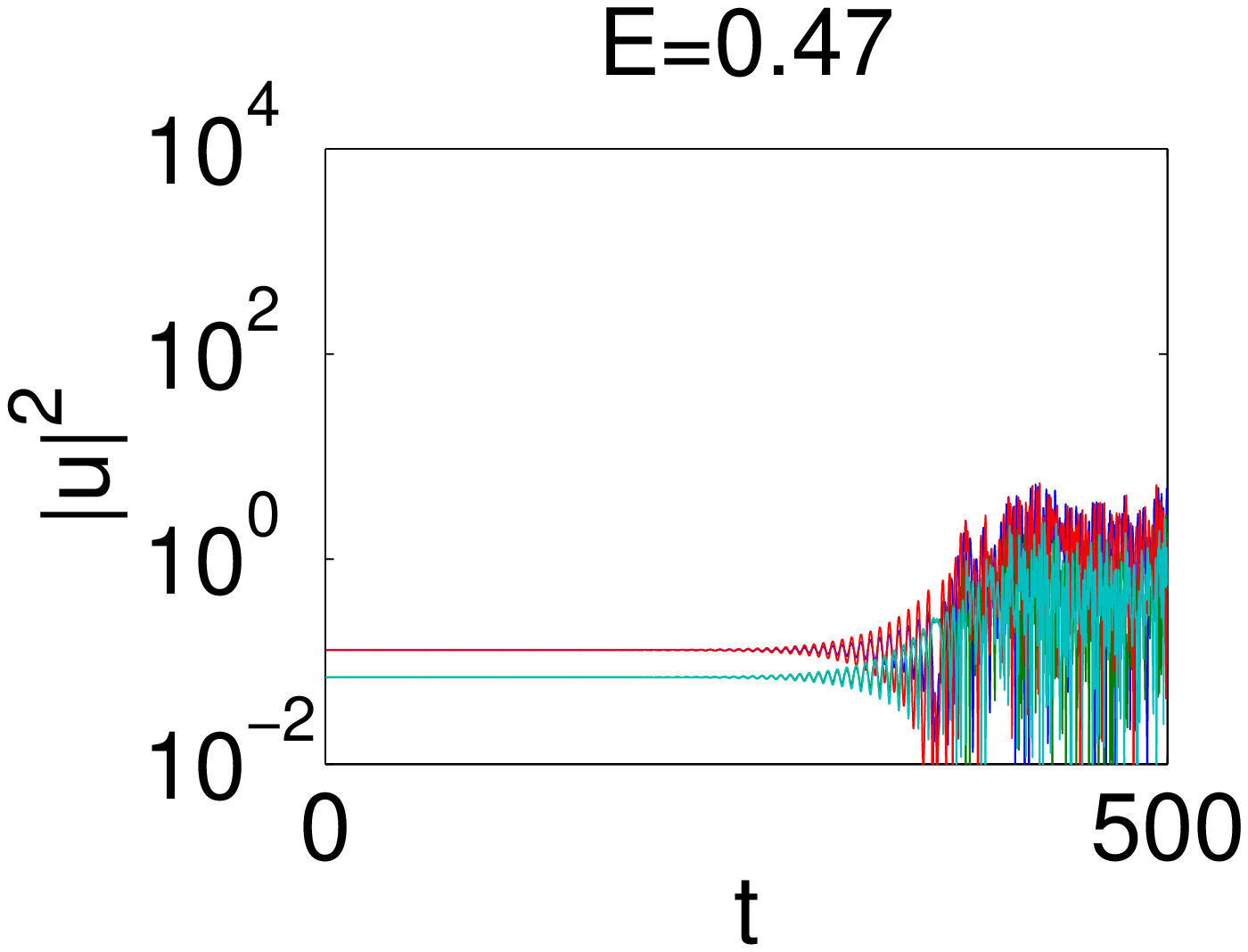}}}%
\subfigure[\ red crosses]{\scalebox{0.28}{\includegraphics{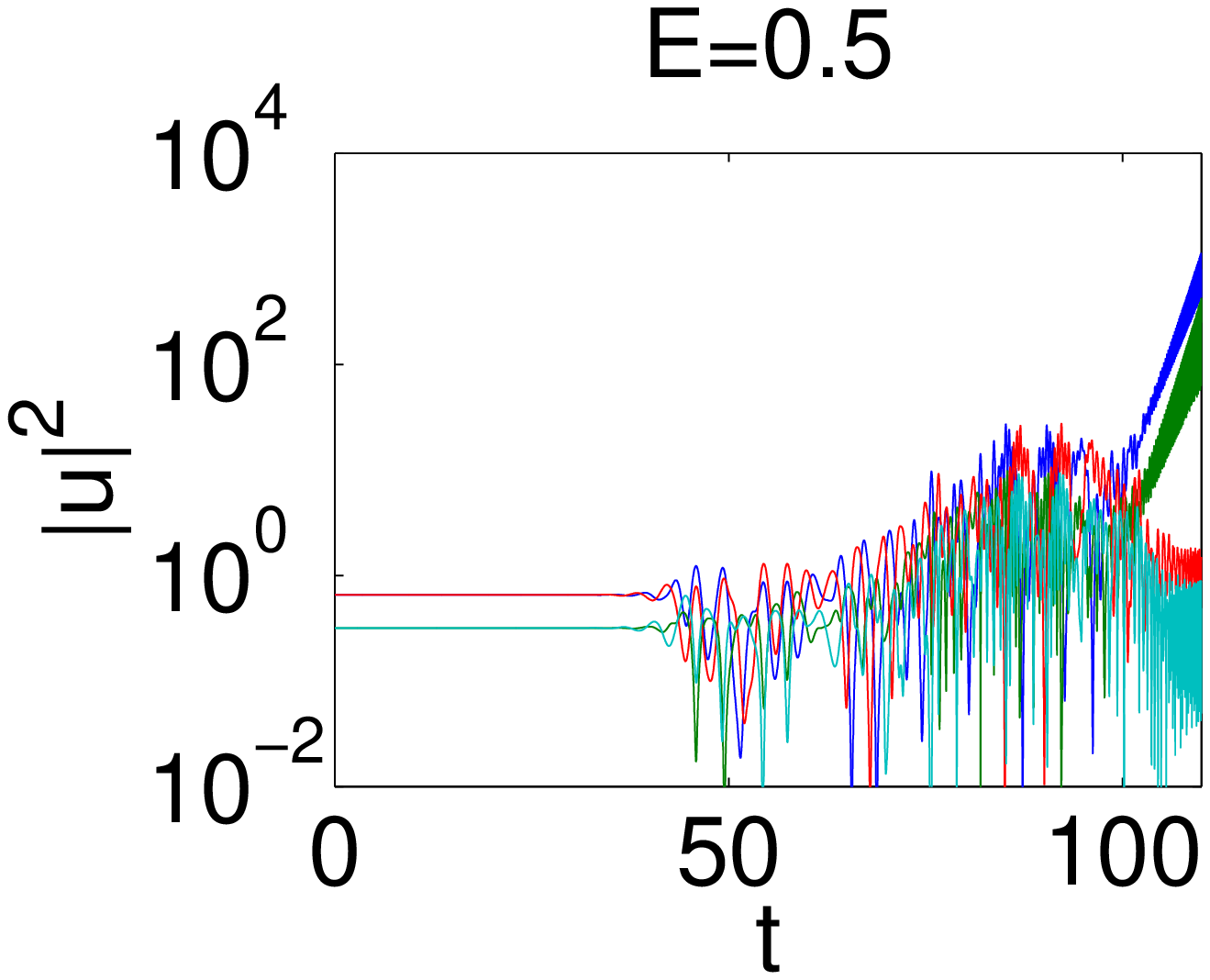}}}\\
\subfigure[\ magenta stars]{\scalebox{0.28}{\includegraphics{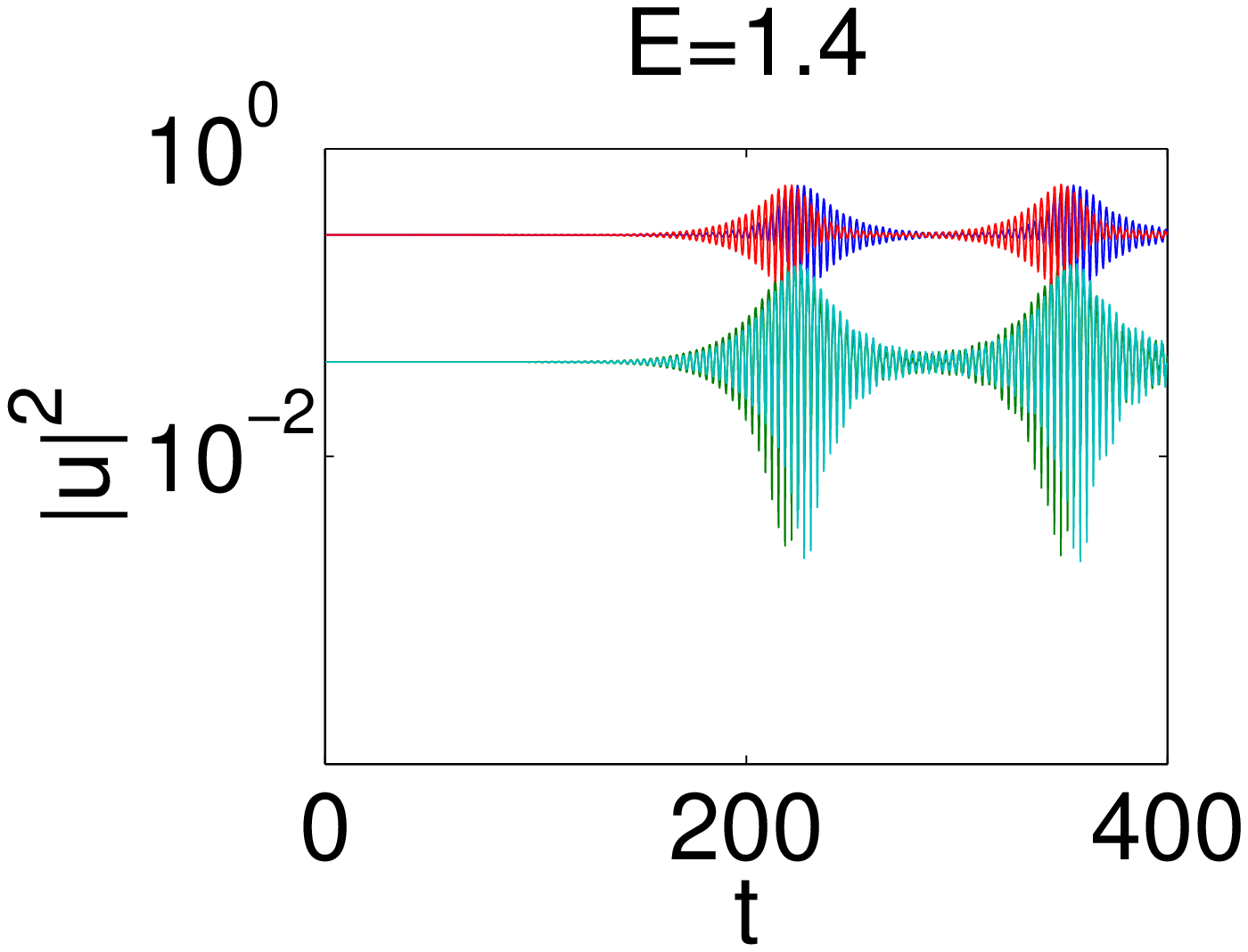}}}%
\subfigure[\ cyan squares]{\scalebox{0.28}{\includegraphics{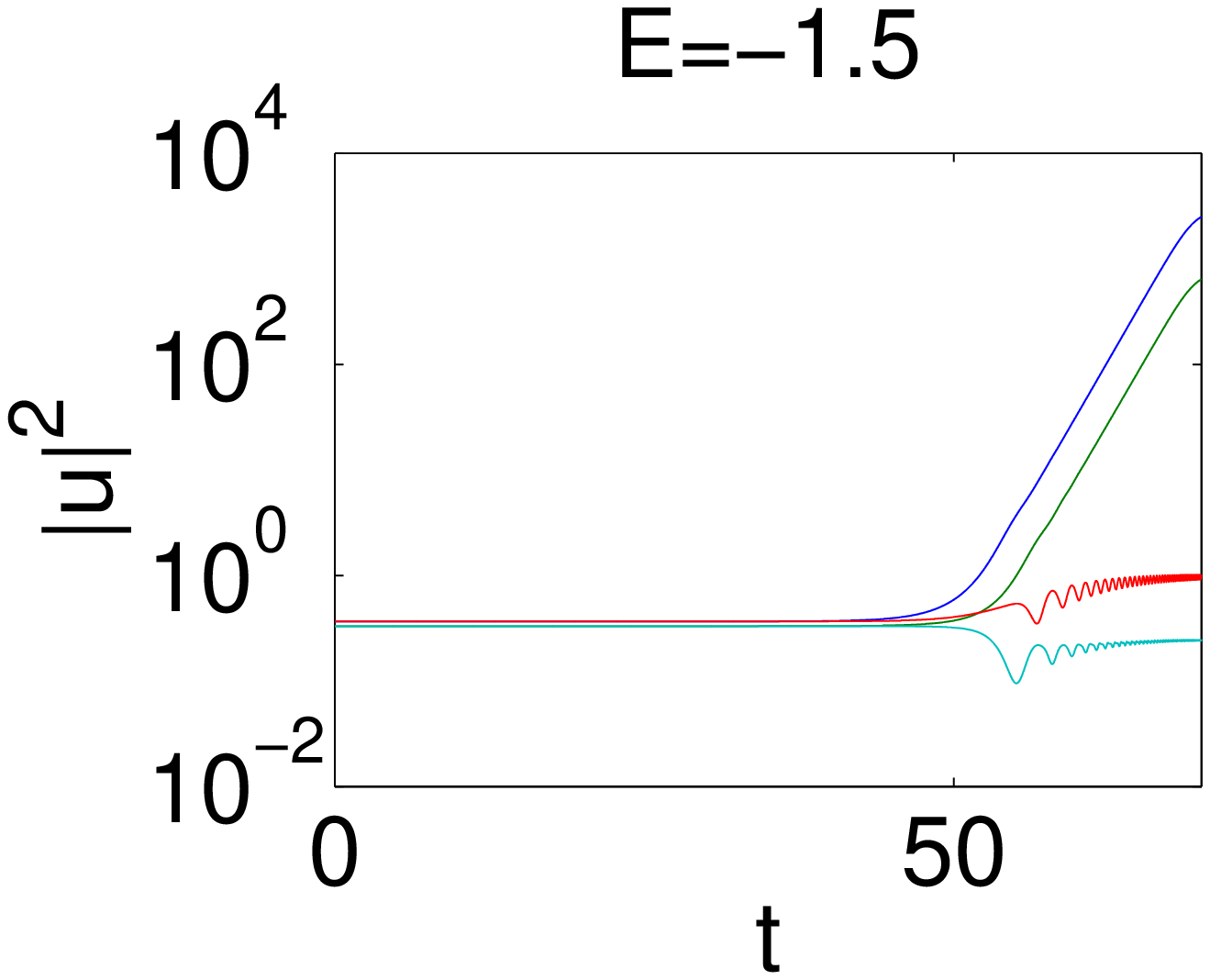}}}\\
\subfigure[\ orange diamonds]{\scalebox{0.28}{\includegraphics{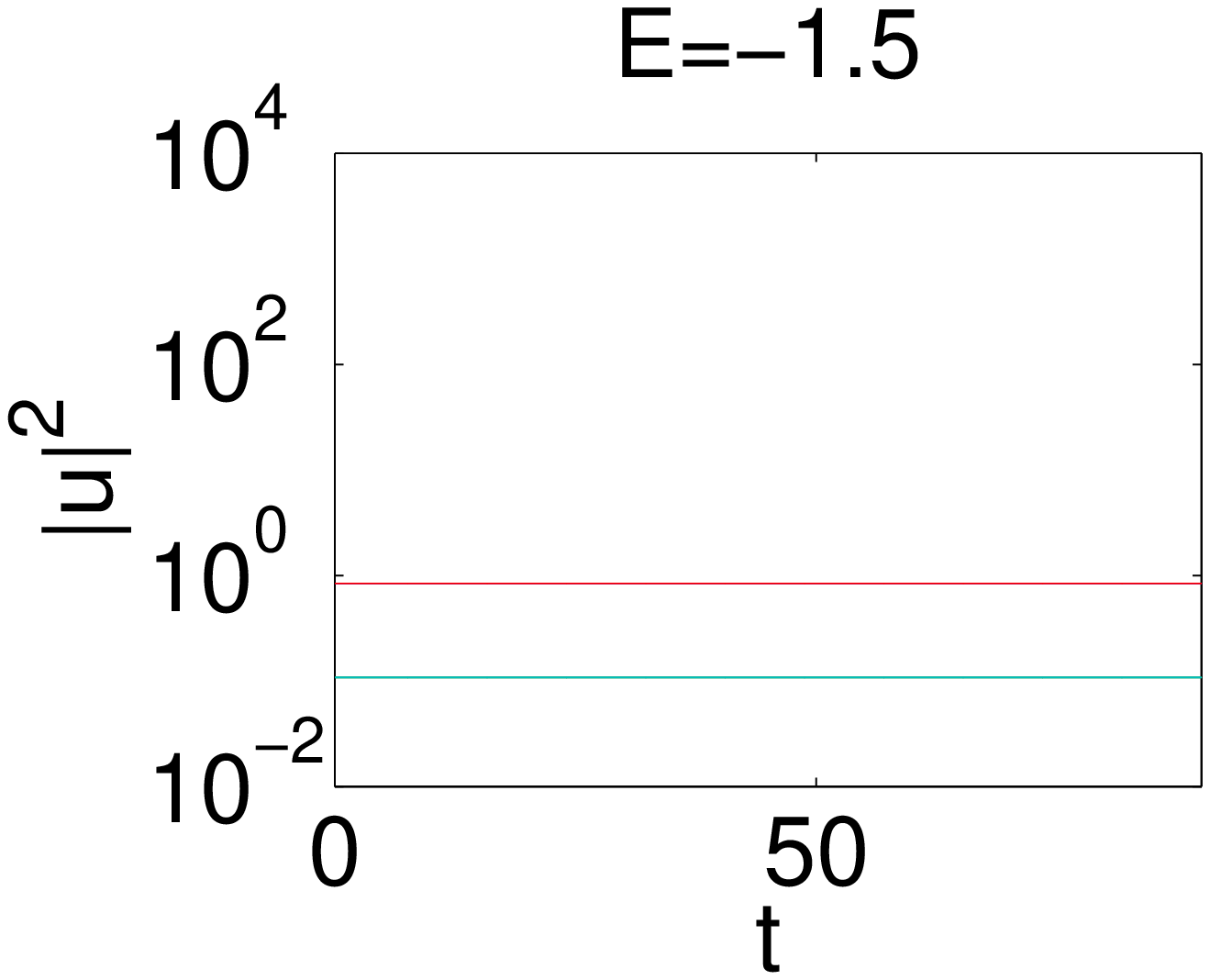}}}%
\subfigure[\ black hexagrams]{\scalebox{0.28}{\includegraphics{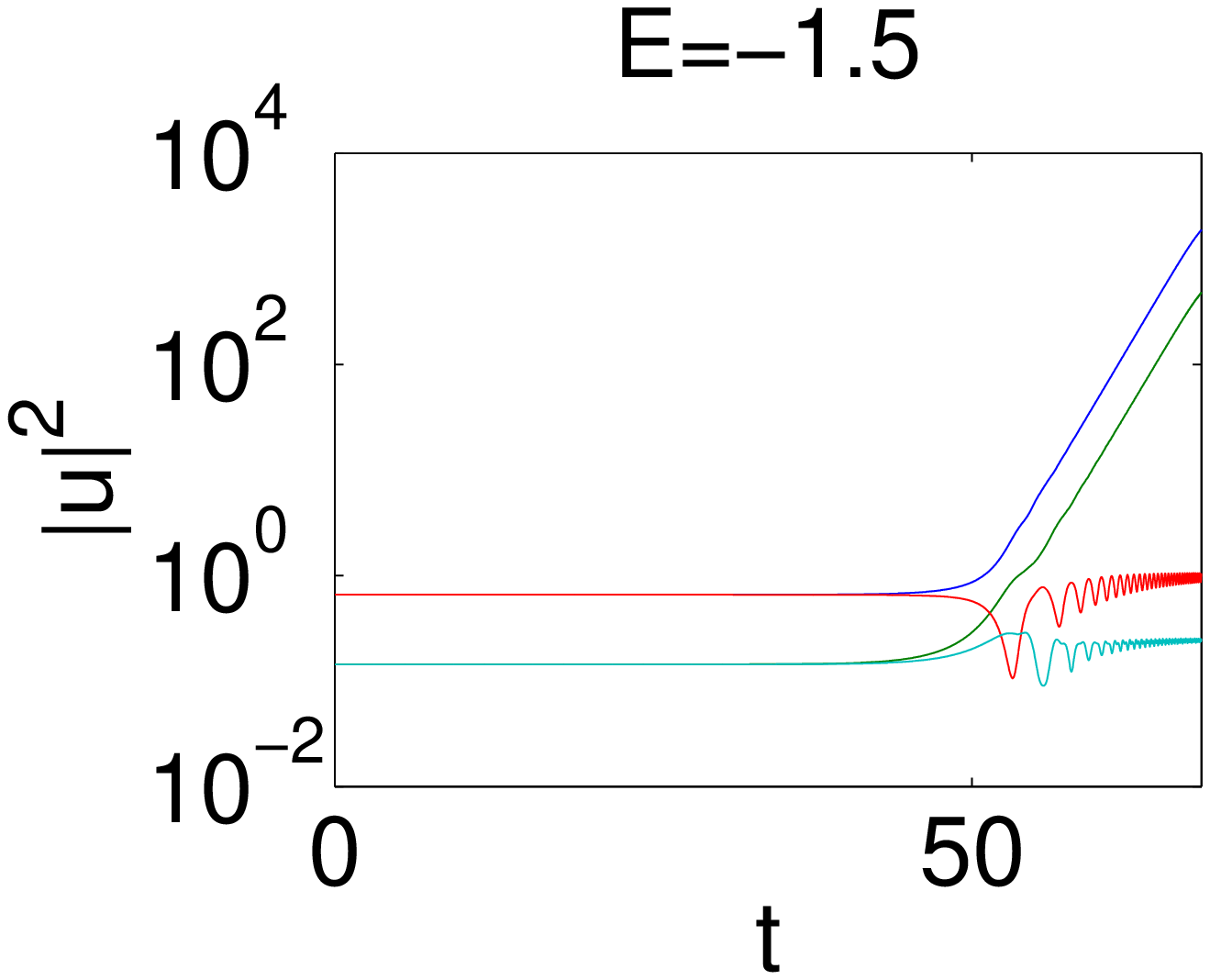}}}\\
\caption{(Color online) Dynamical plots in a semilogarithmic scale
for the $y$-variable (denoting the amplitudes of the fundamental and the
second harmonic) for different  nonlinear modes with $k_1=1$, $k_2 =2$, $q = 0.5$, $\gamma_1=0.1,\ \gamma_2 = 0.5$. The family considered and the value
of the propagation constant are depicted explicitly in each panel.}
\label{chisq_k12q5g15_dyn}
\end{figure}

In Fig.~\ref{chisq_k12q5g15_dyn}, we show the dynamics of the nonlinear modes with $k_1=1$, $k_2 =2$, $q = 0.5$, $\gamma_1=0.1,\ \gamma_2 = 0.5$, which corresponds to Fig.~\ref{chisq_k12q5g15}. We choose different values of $E$ for the different families, usually in order to simulate their
typical unstable behavior under a small perturbation by numerical errors
up to $10^{-7}$ [however, as an exception for the orange diamonds e.g. of
panel (g), we only confirm their generic stability].
In panel (a),(b) and (d), we pick $E=1.5$ for the
blue circles family, $E=0.5$ for the brown pentagrams family and the
red crosses
family, where all of them are unstable. In all three cases here,
the amplitudes of the first waveguide (which features gain)
grow exponentially fast after some
oscillation. The amplitudes of the second waveguide (which sustains loss)
 keep oscillating but also appear to increase in comparison to their
initial values. In panel (c), all the amplitudes of the green pluses family
are relatively constant for a long evolution interval and oscillating
around their initial values, due to its short-living complex quartet of eigenvalues at $E=0.47$. Panel (e) shows the amplitudes of the two waveguides of the
magenta star family which are oscillating quasi-periodically in a similar way at $E=1.4$. Panel (f) and (h) illustrate the instability of cyan squares and black hexagrams where the amplitudes of both harmonics of the first waveguide grow exponentially at about $t=50$ while the amplitudes of the second waveguide do not appear to grow indefinitely
(but contrary to the cubic case, they are also not observed to
systematically decay~\cite{tyugin}). The stable dynamics of the orange
diamonds family at $E=-1.5$ is also plotted in panel (g). Generally, for the
unstable families, we infer either a growth in the first waveguide coupled
with a bounded oscillation in the second waveguide, or a bounded
evolution in both waveguides.

\begin{figure}
\subfigure[\ blue circles]{\scalebox{0.28}{\includegraphics{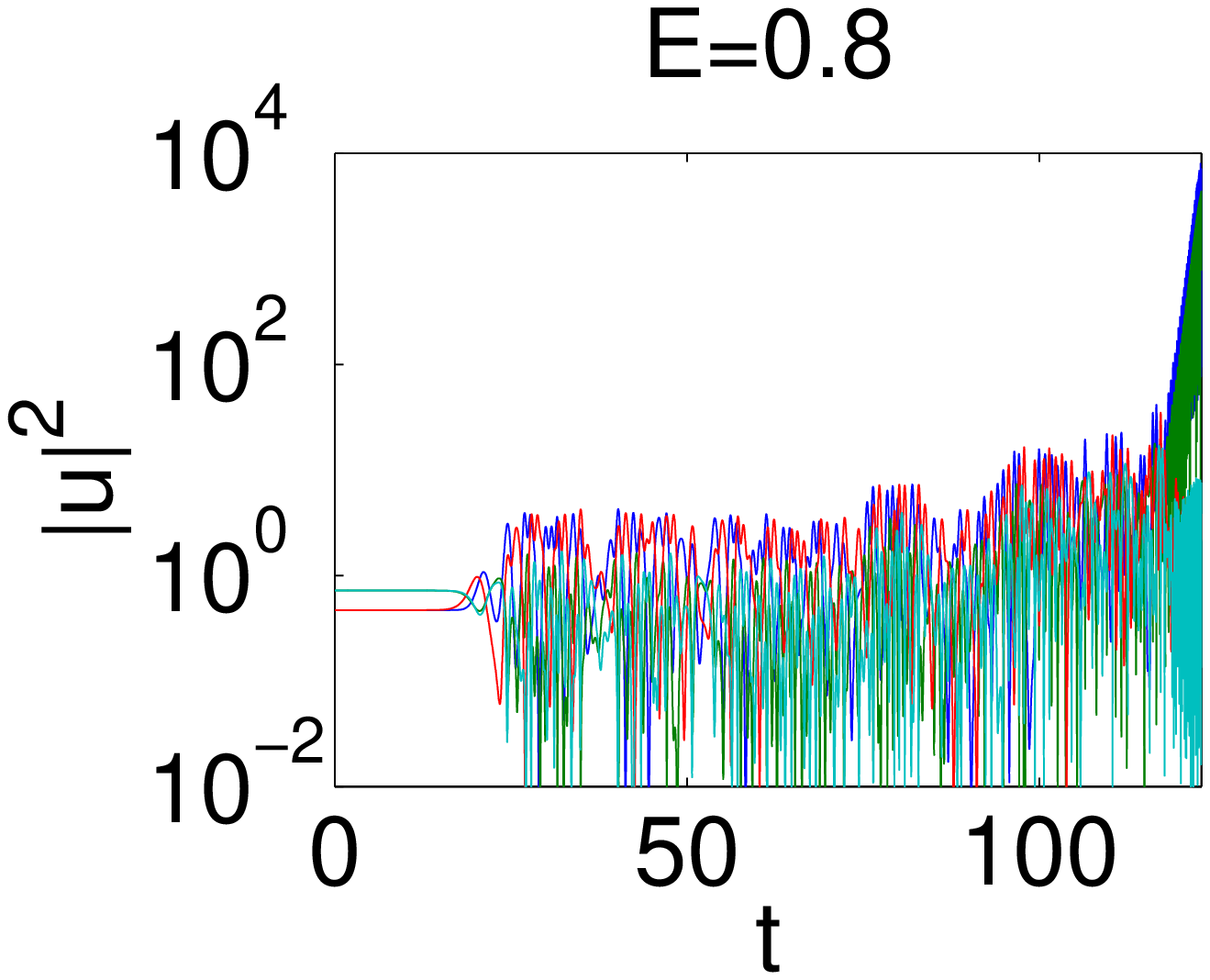}}}%
\subfigure[\ brown pentagrams]{\scalebox{0.28}{\includegraphics{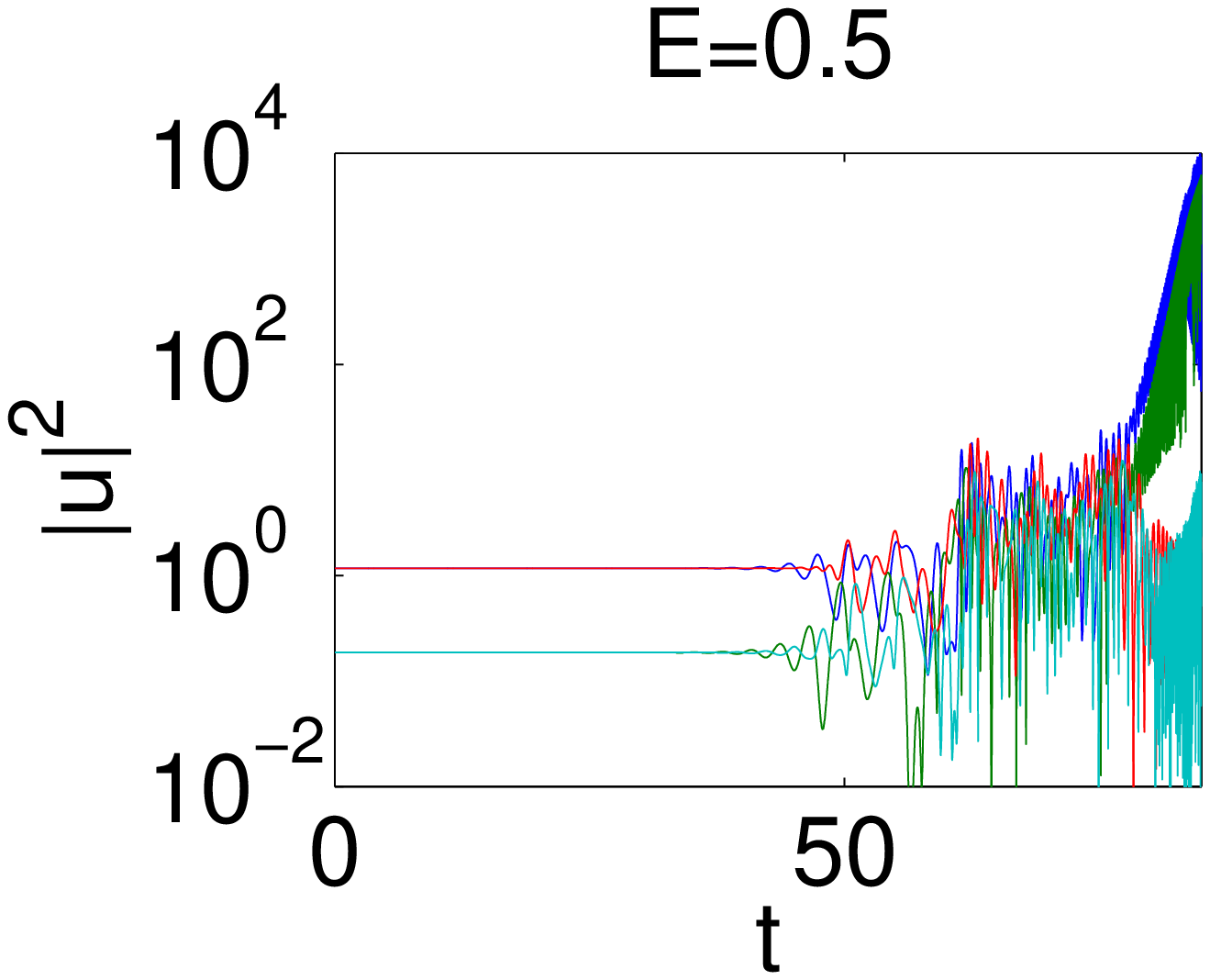}}}\\
\subfigure[\ green pluses]{\scalebox{0.28}{\includegraphics{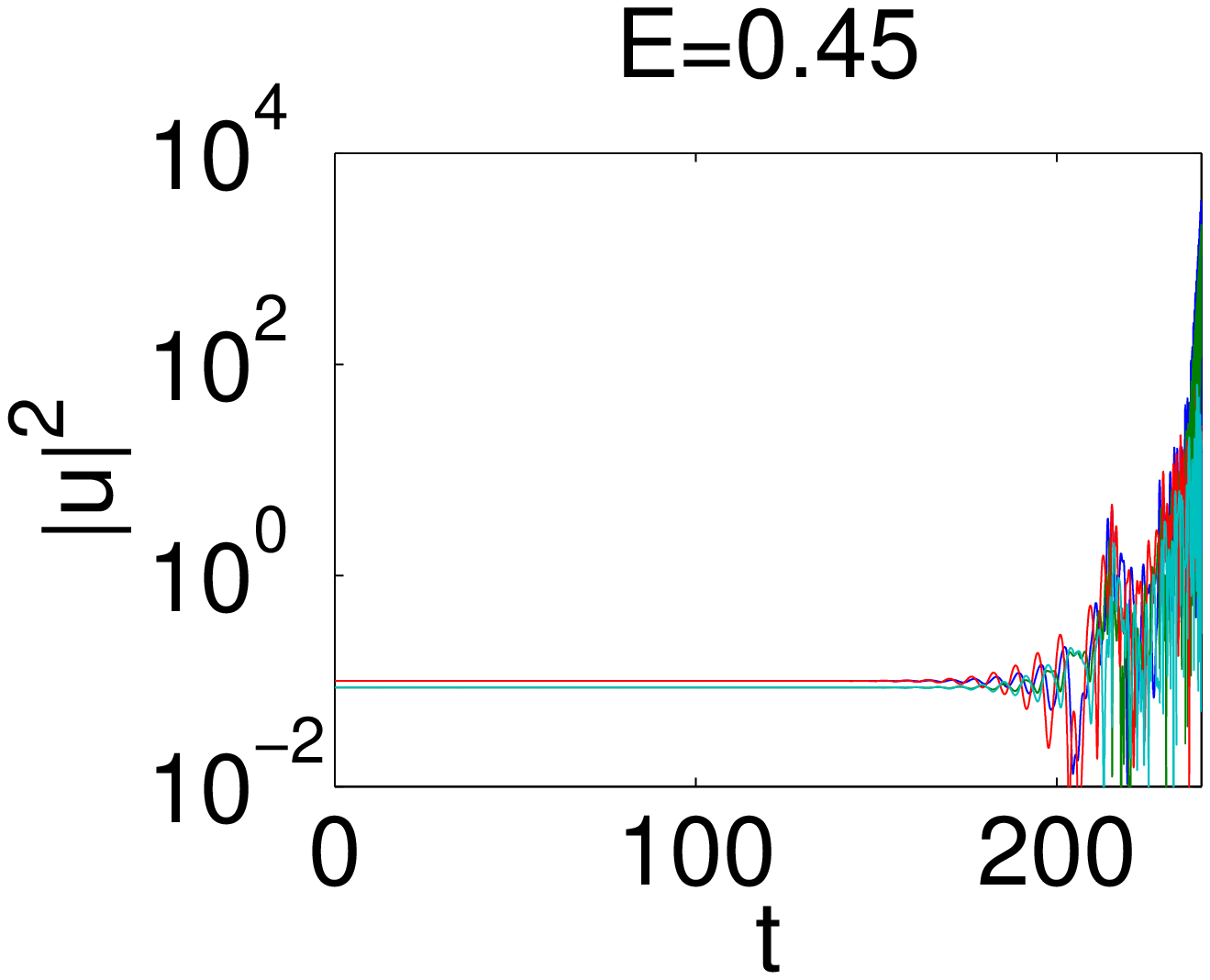}}}%
\subfigure[\ red crosses]{\scalebox{0.28}{\includegraphics{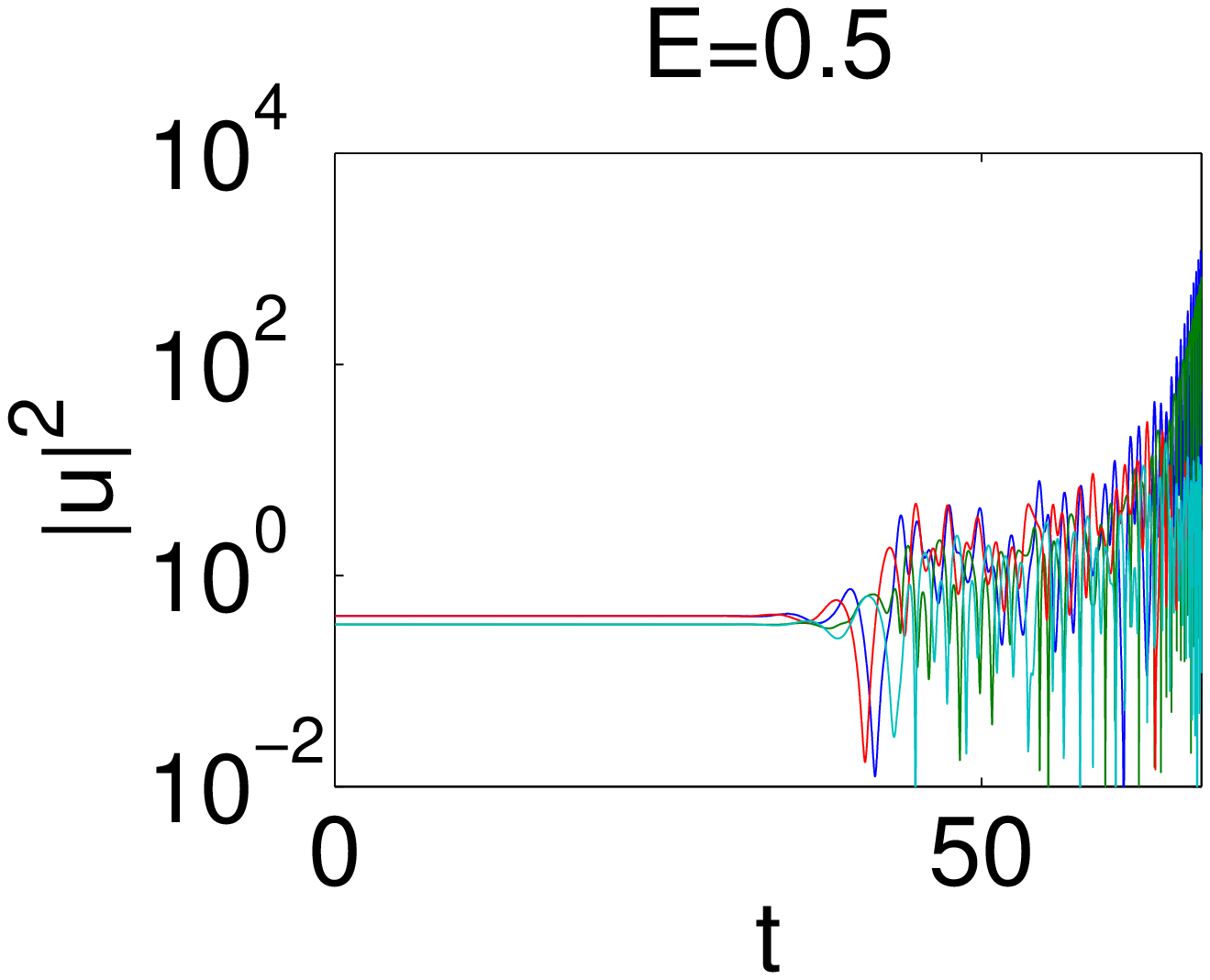}}}\\
\subfigure[\ magenta stars]{\scalebox{0.28}{\includegraphics{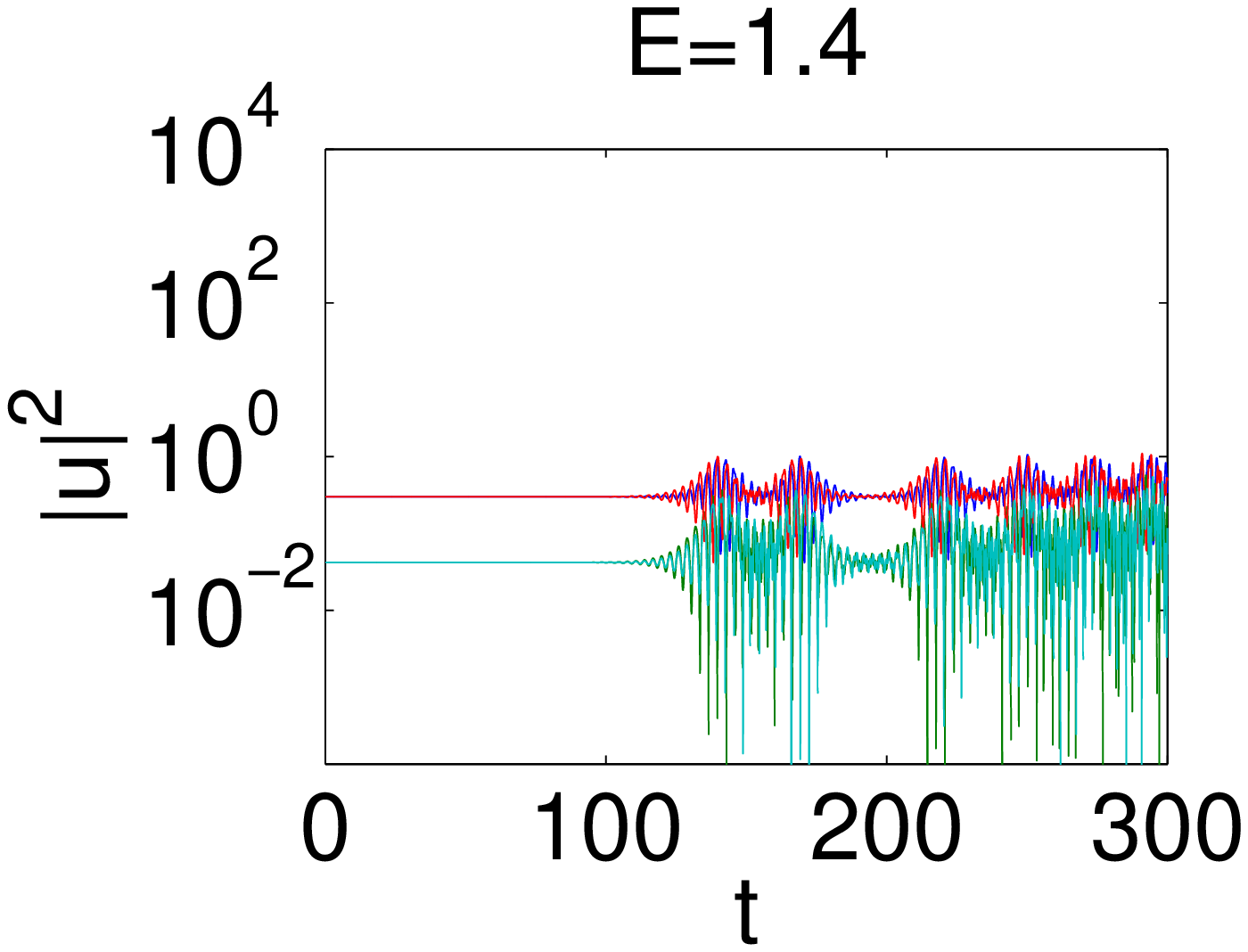}}}%
\subfigure[\ cyan squares]{\scalebox{0.28}{\includegraphics{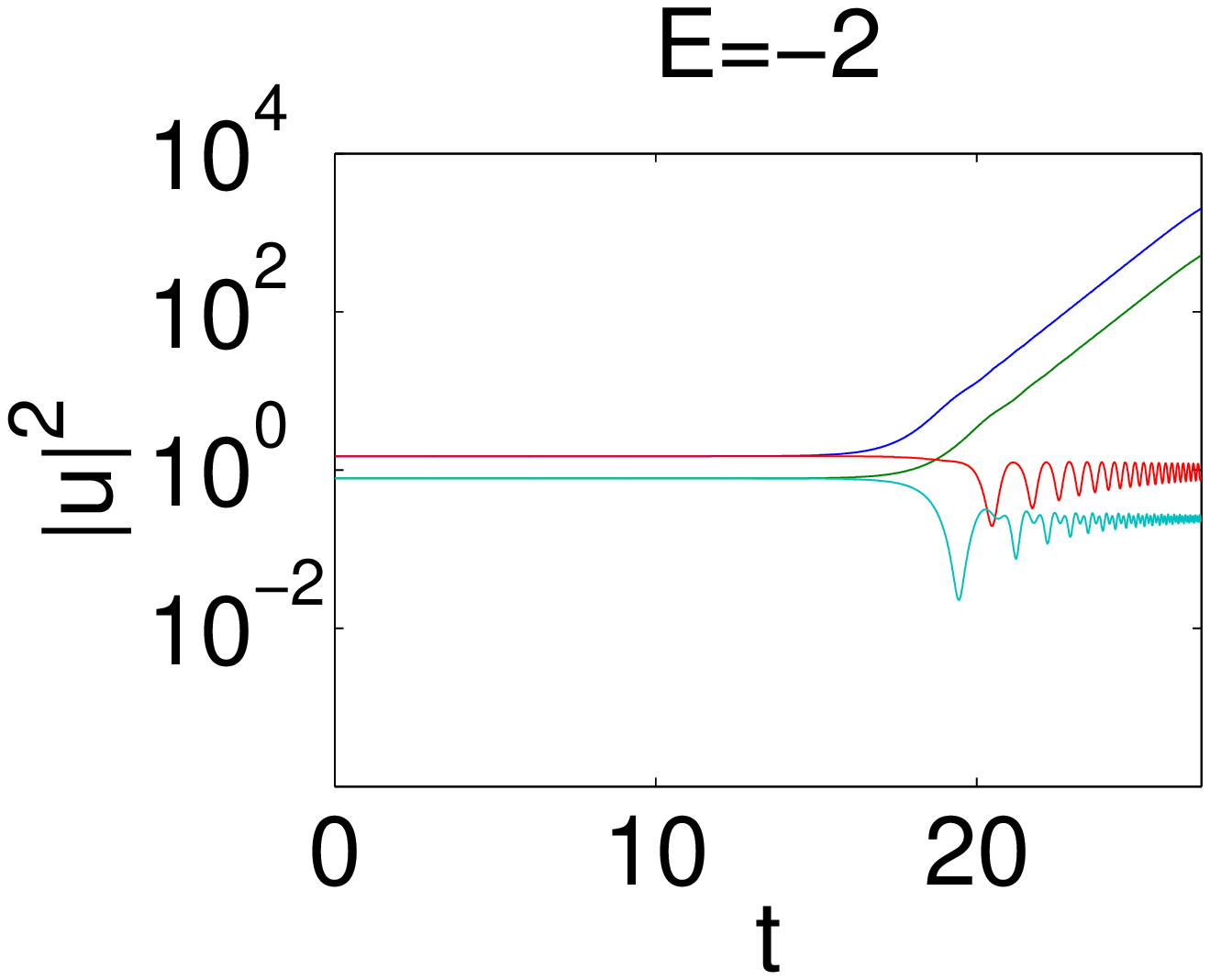}}}\\
\subfigure[\ orange diamonds]{\scalebox{0.28}{\includegraphics{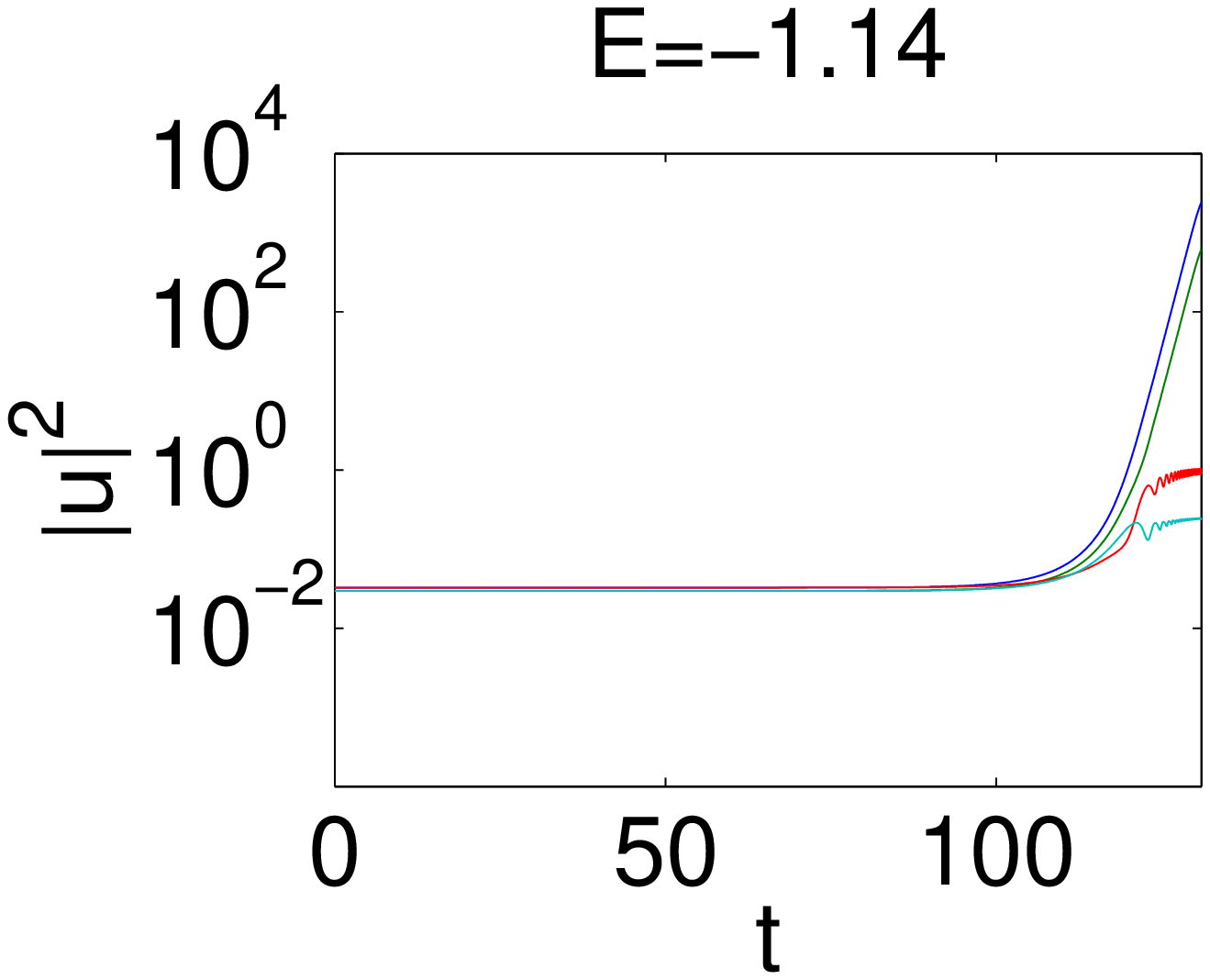}}}%
\caption{(Color online)  Dynamical plots in a semi-logarithmic scale
for the $y$-variable (denoting the amplitudes of the fundamental and the
second harmonic) for different  nonlinear modes with $k_1=1$, $k_2 =2$, $q = 0.5$, $\gamma_1=0.1,\ \gamma_2 = 0.9$.}
\label{chisq_k12q5g19_dyn}
\end{figure}

Fig.~\ref{chisq_k12q5g19_dyn} shows similar dynamic plots corresponding to
the families plotted in Fig.~\ref{chisq_k12q5g19}.
Here all of the blue circles, brown pentagrams, green pluses, and red crosses in panel (a)--(d) are unstable and present similar features as before, with
unbounded growth in the one waveguide (but no decay of amplitude on the
second).
The amplitudes of the magenta stars still oscillate quasi-periodically
around their initial values. In panels (f) and (g), both amplitudes of the
first waveguide grow exponentially. The amplitudes of the second waveguide in cyan squares decay a little and then feature a weak oscillation around their
initial values, whereas for the
orange diamonds family they grow a little and then
feature a similar weak oscillation.

\begin{figure}[tph]
\subfigure[\ blue circles]{\scalebox{0.28}{\includegraphics{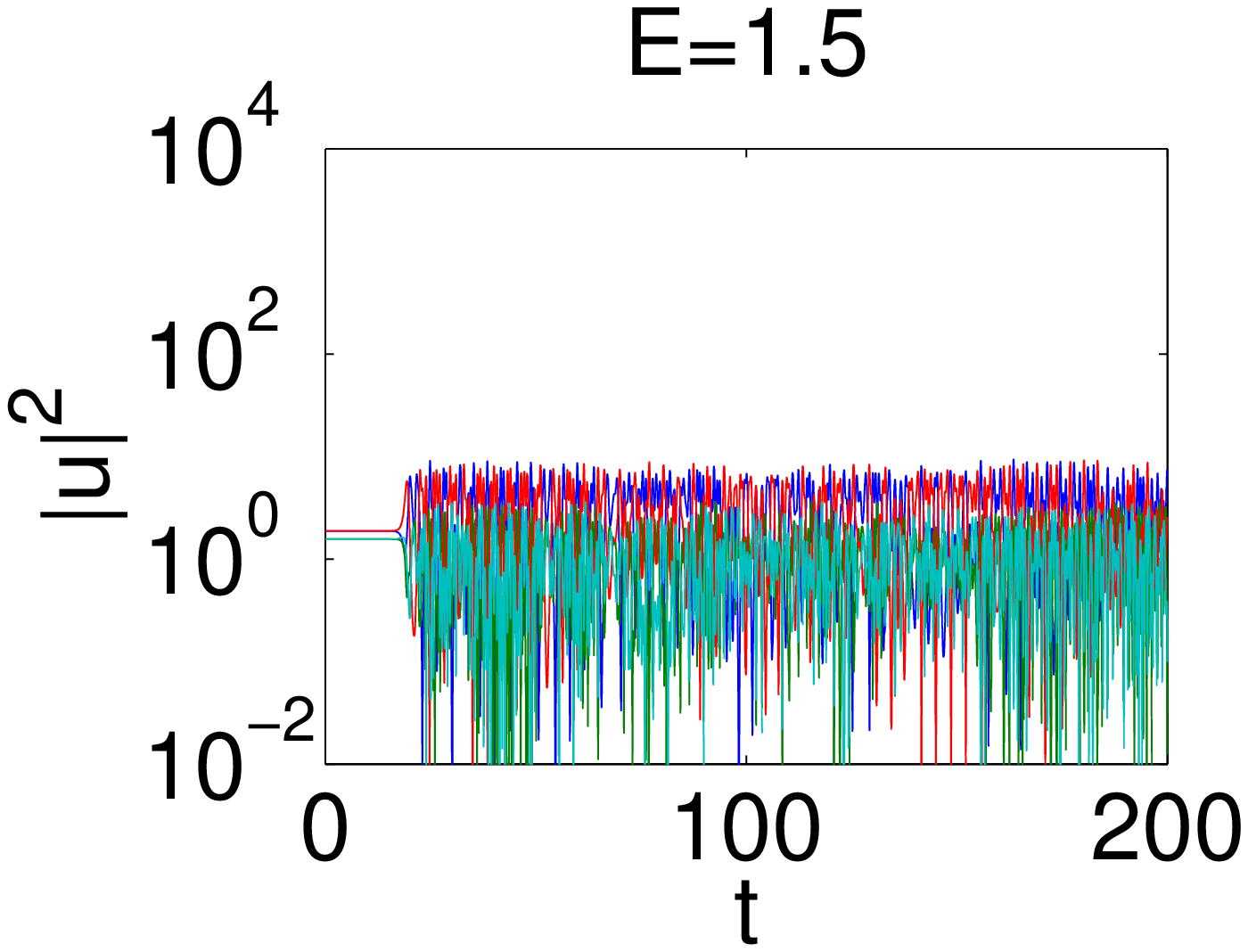}}}%
\subfigure[\ brown pentagrams]{\scalebox{0.28}{\includegraphics{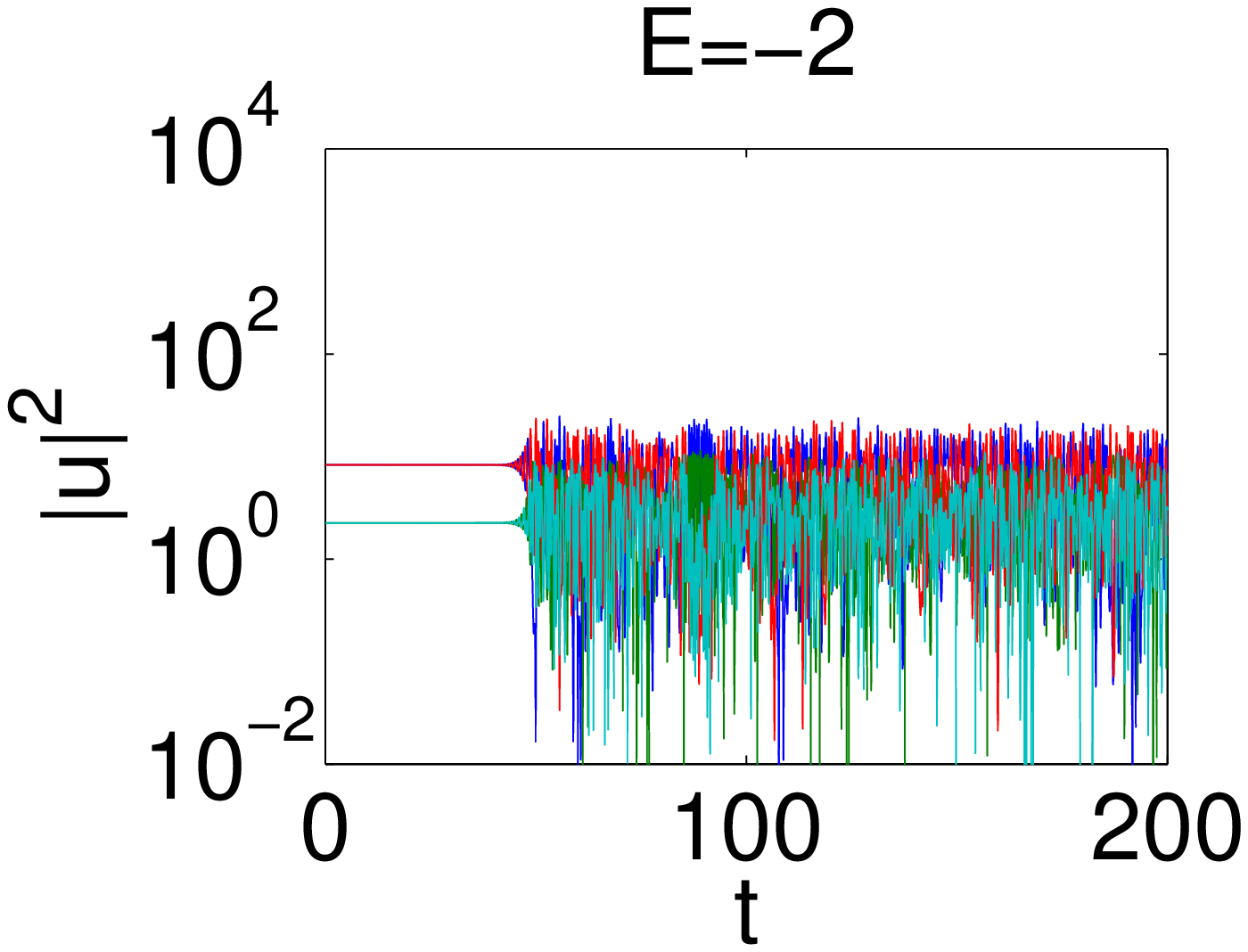}}}\\
\subfigure[\ red crosses]{\scalebox{0.28}{\includegraphics{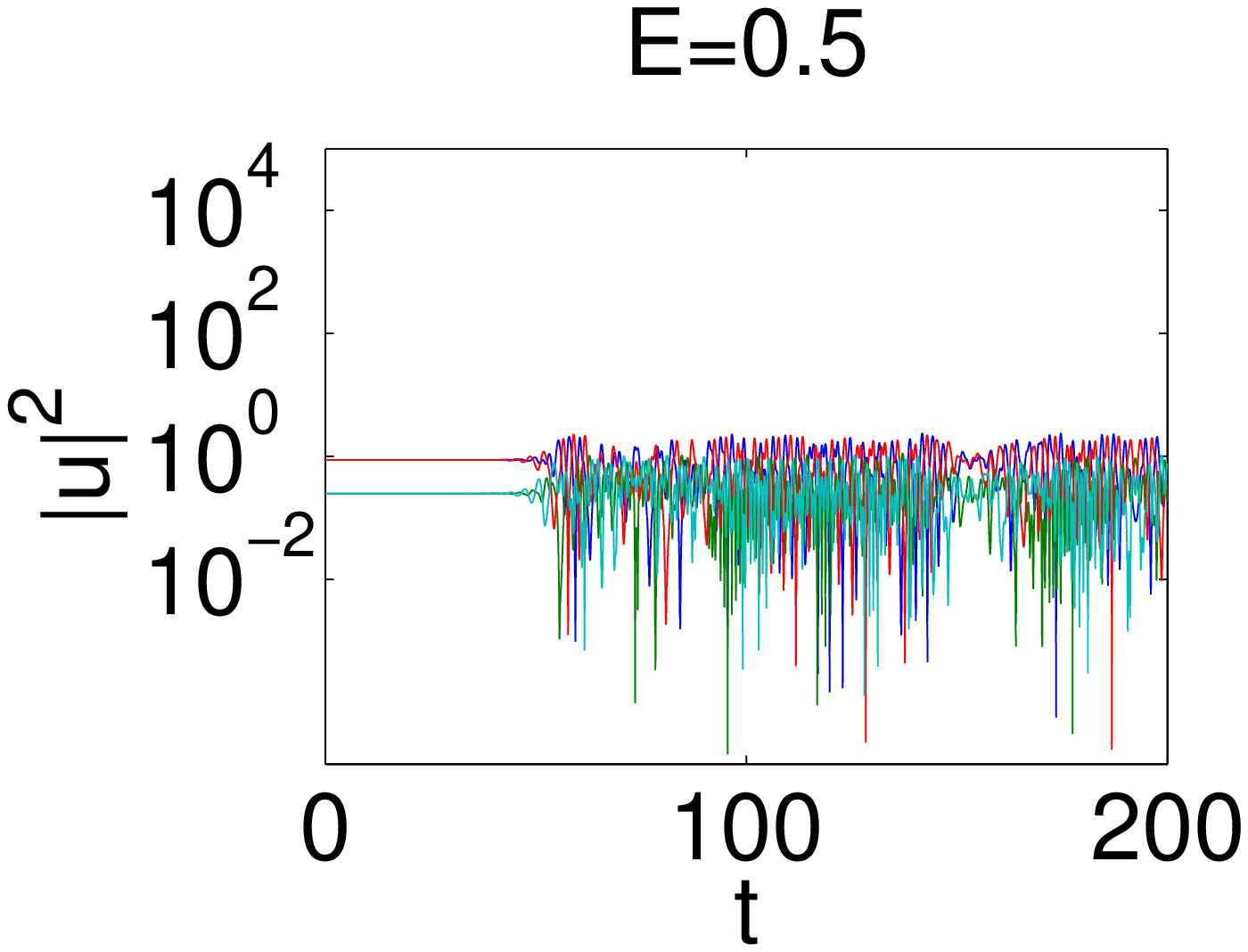}}}%
\subfigure[\ magenta stars]{\scalebox{0.28}{\includegraphics{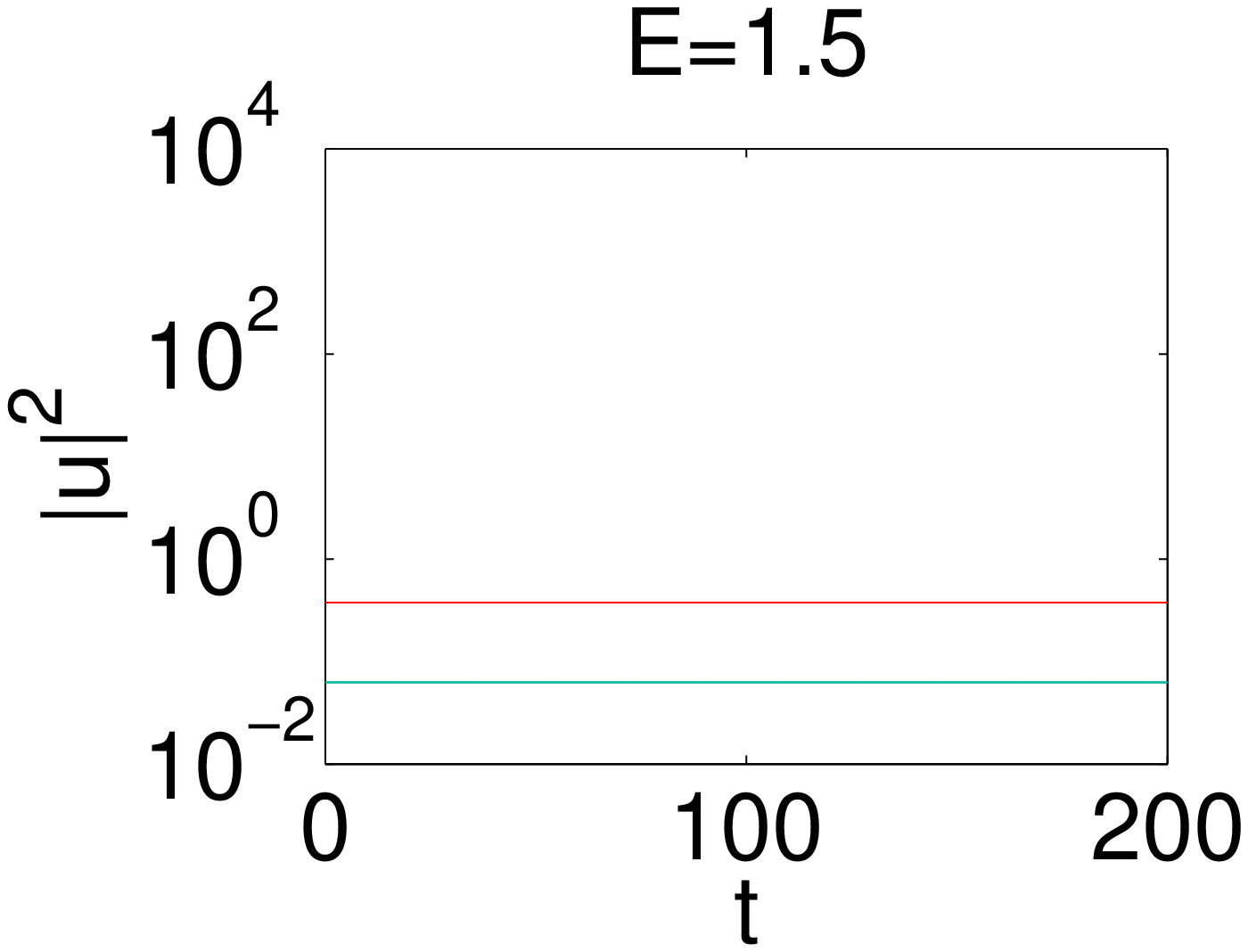}}}\\
\subfigure[\ cyan squares]{\scalebox{0.28}{\includegraphics{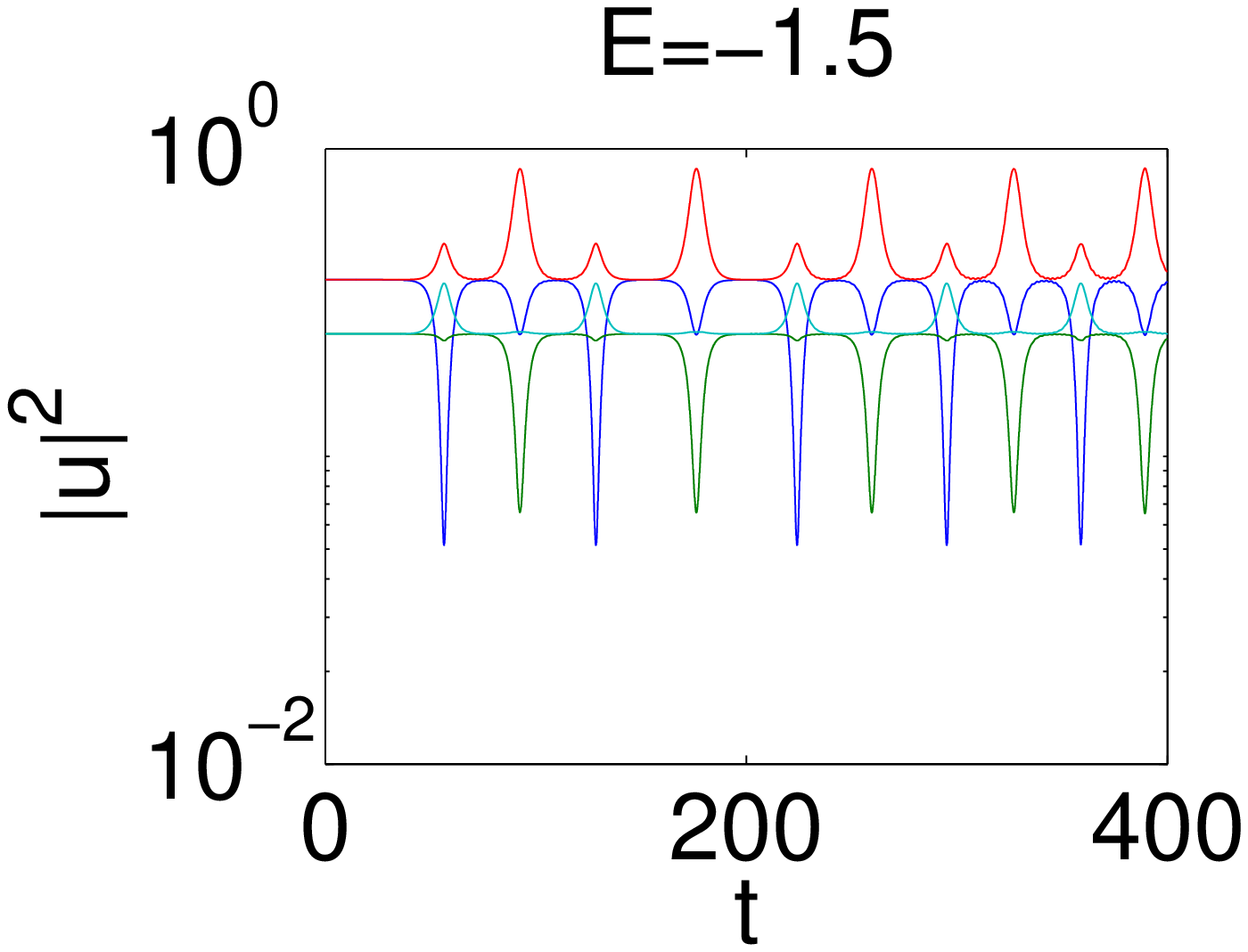}}}%
\subfigure[\ orange diamonds]{\scalebox{0.28}{\includegraphics{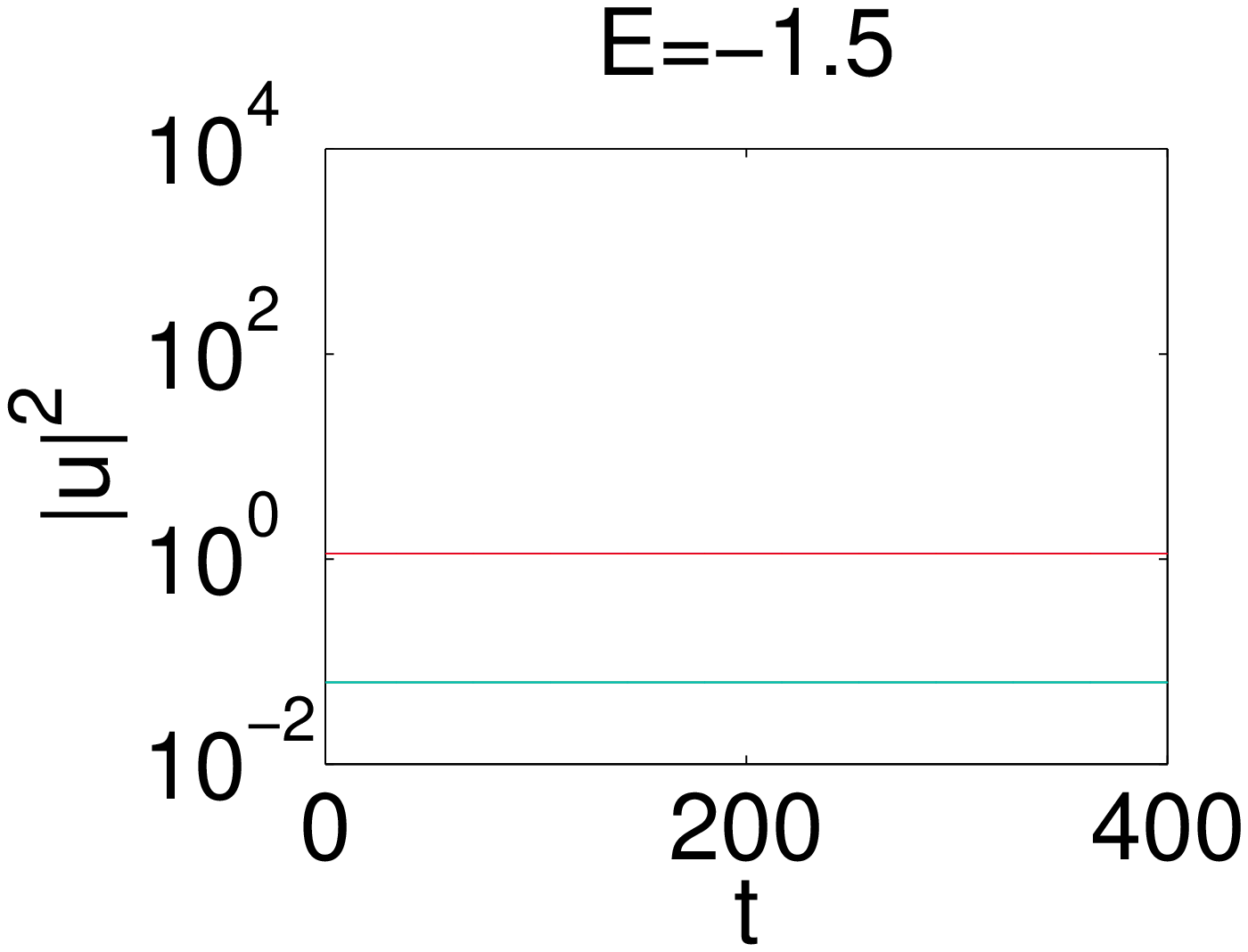}}}
\caption{(Color online)  Dynamical semi-logarithmic plots of nonlinear modes with $k_1=1$, $k_2 =2$, $q = 0.5$, $\gamma_1=0,\ \gamma_2 = 0$.}
\label{chisq_k12q5g00_dyn}
\end{figure}

Fig.~\ref{chisq_k12q5g00_dyn} shows the dynamics of the Hamiltonian
case under the parameters $k_1=1$, $k_2 =2$, $q = 0.5$, $\gamma_1=0,\ \gamma_2 = 0$ that corresponds to Fig.~\ref{chisq_k12q5g00}. Since $\gamma_1=0,\ \gamma_2 = 0$, neither of the two waveguides has a gain or loss profile. Shown in
panels (a)--(c), are all the amplitudes of the harmonics of the the blue
circle, brown pentagram, and red cross families which are oscillating around their initial values, with no trend of indefinite growth or decay, just as expected by the absence
of $\PT$-symmetric terms. In panel (e), the cyan squares family amplitudes now
weakly oscillate
periodically
(but in a way preserving as
they should the Manley-Rowe invariant). Panels (d) and (f) show the stable
dynamics of the magenta stars and the orange diamonds families in this case,
i.e., confirming the dynamical robustness of the families stemming from the
linear limit.

\section{Conclusions and future work}
\label{sec:concl}
In the present work, we considered systematically the features of stationary states of a prototypical $\PT$-symmetric quadratically nonlinear dimer (or coupler).  We  explored different parametric regimes in the two-dimensional plane of gain and loss (for the first and the second harmonic) and in each considered case we identified families of nonlinear modes and addressed the
stability and dynamics of the solutions. We found
 numerous unexpected features that distinguish this system
e.g. from its more well studied sibling, namely the cubic $\PT$-symmetric
nonlinear dimer.

We have started our analysis by considering  the spectrum of the underlying linear problem and found that its eigenvalues always have total energy fully  concentrated either in the first or in the second harmonic of the waveguide.  Turning to the full nonlinear problem, we have established that the found linear solutions give birth to nonlinear modes which (in the vicinity of the bifurcation from linear eigenstate) can be described by means of the small-parameter formal expansions. We have further  revealed  two types of the bifurcations of nonlinear modes from the linear solutions. Namely, the nonlinear modes continued from the linear eigenvectors with total energy concentrated in the first harmonic, have zero  value of the  Manley-Rowe  characteristic at the point of bifurcation. On the other hand, the nonlinear modes arising from the linear eigenvectors with total energy concentrated in the second harmonic bifurcate with finite nonzero  Manley-Rowe  characteristic. Moreover, in the latter case there can exist   two physically distinct  families bifurcating from the same linear state.  These findings were at first quantified by the perturbative formal expansions  which are shown to acquire different forms for the two above-mentioned situations. Then we  confirmed the analytical predictions via  numerical computations of  the
full nonlinear system determining its nonlinear modes.  We have addressed several representative sets of the system parameters and numerically computed continuous families of nonlinear modes as functions of the propagation constant.

Further, we have numerically examined the stability of the identified
families.
Generally, the stability
was found to have rather complex properties but some gross
features could still be discerned such as the systematic robustness
of the modes that emerged from the linear solutions. The dynamics
also features differences from the cubic case, such as the fact
that the lossy waveguide does not typically appear to have a vanishing
amplitude (when growth occurs on the gain side).

Finally, there are numerous directions that one can consider for
future study. On the one hand, one can address simple extensions
of the present dimer, such as the case with  competition
between the signs of $\gamma_1$ and $\gamma_2$ with gain in the
first harmonic but loss in the second (or vice versa). In the way
of extensions to models with more degrees of freedom, one can envision
chains of such dimers at the lattice level (whereby discrete solitary
waves and their properties can be considered)
or even continuum extensions  of the dimer in transverse
continuous directions. In that case, the dimer considered herein
would constitute the limit of homogeneous solutions along such
transverse directions. Some of these possibilities are currently
under investigation and will be reported  in future publications.

\acknowledgments

The work of DAZ and VVK was supported by FCT (Portugal)
through the grants PTDC/FIS-OPT/1918/2012 and
PEst-OE/FIS/UI0618/2011. PGK gratefully acknowledges
support from the US NSF under grant CMMI-1000337,
from the US AFOSR under grant FA9550-12-1-332, and
from the Binational Science Foundation under grant
2010239.

\end{document}